%% file: main.tex
\documentclass{article}
\usepackage{graphicx} 
\usepackage{amsmath,amssymb,amsfonts}
\usepackage{tikz}
\usepackage[backend=biber,style=authoryear]{biblatex}
\usepackage{bm}
\usetikzlibrary{arrows.meta, positioning, decorations.pathreplacing}
\usepackage{bbm}
\usepackage{etoolbox}
\usepackage[ruled,vlined,linesnumbered]{algorithm2e} 
\usepackage{authblk} 
\usepackage[table]{xcolor}
\usepackage{adjustbox}
\usepackage{booktabs}

\title{Risk-Aware Pump Control in Water Supply Systems Using Probabilistic Water Demand and Electricity Price Forecasts}
\author[1]{Jens Kley-Holsteg}

\affil[1]{Chair of Data Science in Energy and Environment, House of Energy, Climate and Finance, University of Duisburg–Essen, Essen, Germany, with corresponding author email. Email: jens.kley-holsteg@stud.uni-due.de}

\begin{document}

\maketitle

\begin{abstract}
Drinking water utilities face uncertain water demand and electricity prices, as well as requirements for reliable and cost-efficient operation. This paper investigates a sequential multi-stage pump scheduling problem under uncertainty for a drinking water supplier participating in the day-ahead auction with subsequent imbalance market settlement. We integrate probabilistic forecasts of water demand and electricity prices into a risk-aware stochastic optimization framework. Reliability and economic risk preferences are represented through a lexicographic objective incorporating Conditional Value-at-Risk and Exceedance Risk measures, where the framework allows to explicitly account for the imbalance market settlement. A numerical study demonstrates that improvements in forecasting performance generally translate into improved policy performance.  Furthermore, we found that the primary benefit of stochastic optimization lies in improved operational robustness rather than lower expected costs. The water demand is found to be a dominant driver of policy quality, the marginal value of further improvements in electricity price forecasts appears comparatively limited, indicating that the economically exploitable component of the price information is already captured to a large extent. Compared with a price-invariant benchmark, the best stochastic policy achieves cost savings of up to $9\%$, indicating that substantial economic benefits can be realized by accepting a carefully controlled increase in imbalance exposure and operational risk.
\end{abstract}

\section{Introduction}
The ongoing transformation of the energy system in Germany is characterized by a growing share of  renewable energy sources (RES) in the electricity mix. 
In recent years this has had a direct influence on the price formation in German electricity markets and hence the cost of electricity consumption. For energy intensive production lines flexibility and predictability became therefore highly valuable properties to cope with the new conditions. 
Due to their high electricity demand, inherent storage flexibility, and the ability to flexibly schedule pump operations, drinking water supply systems have attracted considerable research interest in the context of optimization-based operational cost minimization, as demonstrated for example by \textcite{Singh2020}.
\\
This transformation introduces new operational challenges for water suppliers, as economic objectives such as cost efficiency must be balanced simultaneously with resilience and sustainability considerations. These objectives must be pursued while ensuring the primary responsibility of water utilities, namely the reliable provision of sufficient, affordable, and high-quality drinking water to private and public consumers.

To address these operational challenges, a broad range of optimal and real-time control approaches have been proposed, as discussed by \textcite{Malajetmarova2017}. Among these, the Model Predictive Control (MPC) framework has attracted considerable attention due to its capability to explicitly incorporate forecasts and system dynamics into sequential real-time decision-making, as outlined by \textcite{Castelletti2023}. In parallel, Dynamic Programming approaches and their approximations, including reinforcement learning, have been widely applied to formulate and solve sequential decision problems over time, as demonstrated, for example, by \textcite{Candelieri2018} and \textcite{Hajgato2020}. To solve the resulting optimization problems, mixed-integer linear programming (MILP) and mixed-integer non-linear programming (MINLP) formulations are frequently employed, depending on the required level of hydraulic detail and the adopted approximation strategies, as shown by \textcite{Shao2024}. In addition, heuristic and metaheuristic solution approaches, particularly genetic and evolutionary algorithms, have been proposed to address highly non-linear and computationally complex optimization problems, as summarized by \textcite{Parvaze2023}. Furthermore, competing operational requirements have strengthened the relevance of multi-objective optimization approaches within water system operation, as demonstrated by \textcite{Musabandesu2025} and \textcite{Kidanu2023}.
\\
Substantial attention has been devoted to the management of operational risks and reliability of water distribution systems under uncertainty \parencite{Mkireb2019, Guo2020, Zhou2024}. In this context, water demand and electricity price formation constitute core exogenous processes whose forecasts propagate uncertainty into optimization frameworks. Both forecasting fields have been studied extensively in the literature \parencite{Donkor2014,Nowotarski2018}.

Against this background, the present work develops a methodology for integrating probabilistic forecasts into risk-aware stochastic optimization and investigates the resulting impact on operational decision-making under uncertainty. The main contributions of this work are as follows:

\begin{itemize}
    \item Integration of probabilistic forecasting models for water demand and electricity prices into a sequential stochastic optimization framework for optimal pump control.

    \item The explicit consideration of risk preferences regarding system reliability and economic cost efficiency through the introduction of a risk aware lexicographic objective function.
    
    \item The explicit modeling of the imbalance market, allowing the optimization framework to anticipate imbalance-related risks and opportunities based on forecast information.
\end{itemize}

The remainder is structured as follows. First, a generalized optimal control problem for a simplified water distribution network is introduced. The associated stochastic optimization problem is then formulated within a sequential decision-making framework. Building on this, the specific forecasting methodologies and operational policy models are presented to obtain a tractable solution approach. Finally, the proposed policies with their corresponding forecasting models are evaluated within a numerical case study, followed by a discussion of the obtained results.

\subsection{Problem statement}
\label{sec:problem statement}
The methodology is evaluated using operational data from a real-world drinking water supplier. To emphasize the methodological aspects of forecast-informed stochastic optimization, the underlying waterworks is represented by a generic operational decision problem comprising fixed-flow pumps, a storage facility, and a district metering area. The operator seeks to ensure a reliable water supply at minimal cost, procuring electricity on wholesale markets and settling deviations between procured and realized consumption through imbalance mechanisms. The following assumptions are adopted:
\\

\textbf{Assumption 1.} \label{ass:systemhydraulics} \textit{Hydraulic system.}
The hydraulic system is deliberately simplified to retain only the operational relationships that are common to most drinking water distribution systems, namely the interactions between water demand, storage, pump operation, electricity procurement, and imbalance market settlement. This abstraction improves the transferability of the proposed framework by avoiding site-specific hydraulic details that differ across water suppliers. At the same time, it reduces model complexity, allowing the study to focus on the treatment of forecast uncertainty within the optimization problem. The framework is readily extensible, such that network-specific hydraulic models and operational constraints can be incorporated without modifying the underlying forecasting and optimization concepts. Specifically, we assumed that all pumped water passes through a storage facility before being supplied to end customers, allowing storage dynamics to be represented by a simple mass balance. Furthermore, friction losses between the waterworks and storage facility are neglected, such that pumping head depends solely on the net elevation difference between the waterworks and the storage water level. Finally, ramping times of fixed-speed pumps are assumed negligible, implying binary operating states.
\\

\textbf{Assumption 2} \label{ass:electricityprocurment}.
\textit{Electricity procurement.}
Electricity is procured exclusively through the EPEX SPOT day-ahead auction (DA), where procurement decisions must be submitted by 12:00 CET/CEST on day \(d\) for delivery on day \(d+1\). The realized price vector is published around 12:00 CET/CEST on day \(d\). Deviations between procured electricity and realized consumption are settled through the imbalance market, which serves to stabilize the power system by compensating residual imbalances. The corresponding imbalance settlement price (IBP) is determined ex post based on the realized system balancing situation. Owing to its symmetric structure, the IBP may result in either imbalance costs or imbalance revenues depending on the market position of the market participants. Although balancing responsible parties are generally obliged to maintain balanced positions within their balancing groups, incentives exists to deviate from the DA commitment \parencite{Bnetza2022}. Wile preliminary IBP estimates are published shortly after delivery, the final imbalance prices become available only with a substantial delay. Long-term hedging instruments, participation in additional short-term electricity markets, and additional operational cost components are neglected.
\\

\begin{figure}[ht]
\centering
\begin{tikzpicture}

\def\W{12}

\def\wA{2.4}   
\def\wB{2.0}   
\def\wC{3.2}   
\def\wD{2.0}   
\def\wE{2.4}   

\def\H{1}

\fill[red!80] (0,0) rectangle (\wA,\H);
\fill[orange!70] (\wA,0) rectangle ({\wA+\wB},\H);
\fill[green!60!black!40] ({\wA+\wB},0) rectangle ({\wA+\wB+\wC},\H);
\fill[orange!70] ({\wA+\wB+\wC},0) rectangle ({\wA+\wB+\wC+\wD},\H);
\fill[red!80] ({\wA+\wB+\wC+\wD},0) rectangle (\W,\H);

\node at ({\wA/2}, {0.5*\H}) {Emergency};
\node at ({\wA+\wB/2}, {0.5*\H}) {Critical};
\node at ({\wA+\wB+\wC/2}, {0.5*\H}) {Admissible};
\node at ({\wA+\wB+\wC+\wD/2}, {0.5*\H}) {Critical};
\node at ({\wA+\wB+\wC+\wD+\wE/2}, {0.5*\H}) {Emergency};

\draw (0,0) -- (0,-0.3);
\draw (\wA,0) -- (\wA,-0.3);
\draw ({\wA+\wB},0) -- ({\wA+\wB},-0.3);
\draw ({\wA+\wB+\wC},0) -- ({\wA+\wB+\wC},-0.3);
\draw ({\wA+\wB+\wC+\wD},0) -- ({\wA+\wB+\wC+\wD},-0.3);
\draw (\W,0) -- (\W,-0.3);

\node[below] at (0,-0.3) {};
\node[below] at (\wA,-0.3) {$lbc$};
\node[below] at ({\wA+\wB},-0.3) {$lba$};
\node[below] at ({\wA+\wB+\wC},-0.3) {$uba$};
\node[below] at ({\wA+\wB+\wC+\wD},-0.3) {$ubc$};
\node[below] at (\W,-0.3) {};

\end{tikzpicture}
\caption{Operating ranges for storage and pump control with their corresponding control-relevant bounds: lower bound critical (lbc), lower bound admissible (lba), upper bound admissible (uba), and upper bound critical (ubc).}
\label{fig:operating_ranges}
\end{figure}
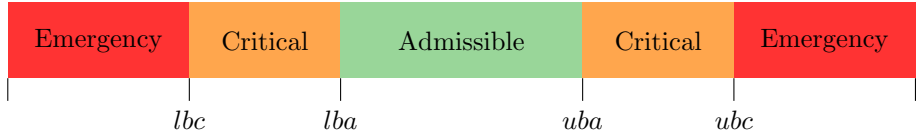

\textbf{Assumption 4.} \label{ass:uncertaintyandrisk}
\textit{Uncertainty modeling and risk preferences.}
Uncertainty is assumed to enter the system exclusively through water demand (WD), day-ahead prices (DAP) and IBP. Although there are additional sources of uncertainty in water distribution systems, such as model misspecification and data assimilation errors \parencite{Hutton2014}, all remaining components of the system are assumed to be deterministic. The operator is assumed to exhibit risk preferences along two dimensions. First, with respect to \emph{system reliability}, adverse demand realizations can induce violations of storage and pump operating constraints. To capture different levels of operational severity, admissible, critical, and emergency operating regimes are distinguished (Figure~\ref{fig:operating_ranges}). Although operation within the admissible range allows purely economic optimization, violations of critical and emergency ranges receive increasing priority over economic objectives. Second, with respect to \emph{economic performance}, the operator seeks to limit exposure to extreme cost realizations. Since alternative hedging instruments are not considered, economic risk management is restricted to demand response within the day-ahead and the allocation of positions between the day-ahead and imbalance markets.
\\

\textbf{Assumption 5.} \label{ass:studydesign}
\textit{Study design.}
To evaluate the proposed policy and corresponding forecasting models, a computationally tractable study design is introduced. Specifically, we sample optimization runs from the test period and solve them independently. Unlike classical MPC, the proposed approach omits re-optimization during policy execution. Nevertheless, the resulting policies remain state-dependent and can adapt to newly observed information. This design affects both initialization and termination: since no preceding optimization run exists, the operator is assumed to be able to retroactively participate in the day-ahead market on day \(d-1\), implying that day-ahead prices for \(t<12\) are known and deterministic. Likewise, the absence of a subsequent optimization run induces a finite horizon and potential end-of-horizon effects.
\\

\subsection{Data Description and data preprocessing}
\label{sec:preprocessing} 
\begin{figure}[ht]
    \centering
    \includegraphics[width=\textwidth]{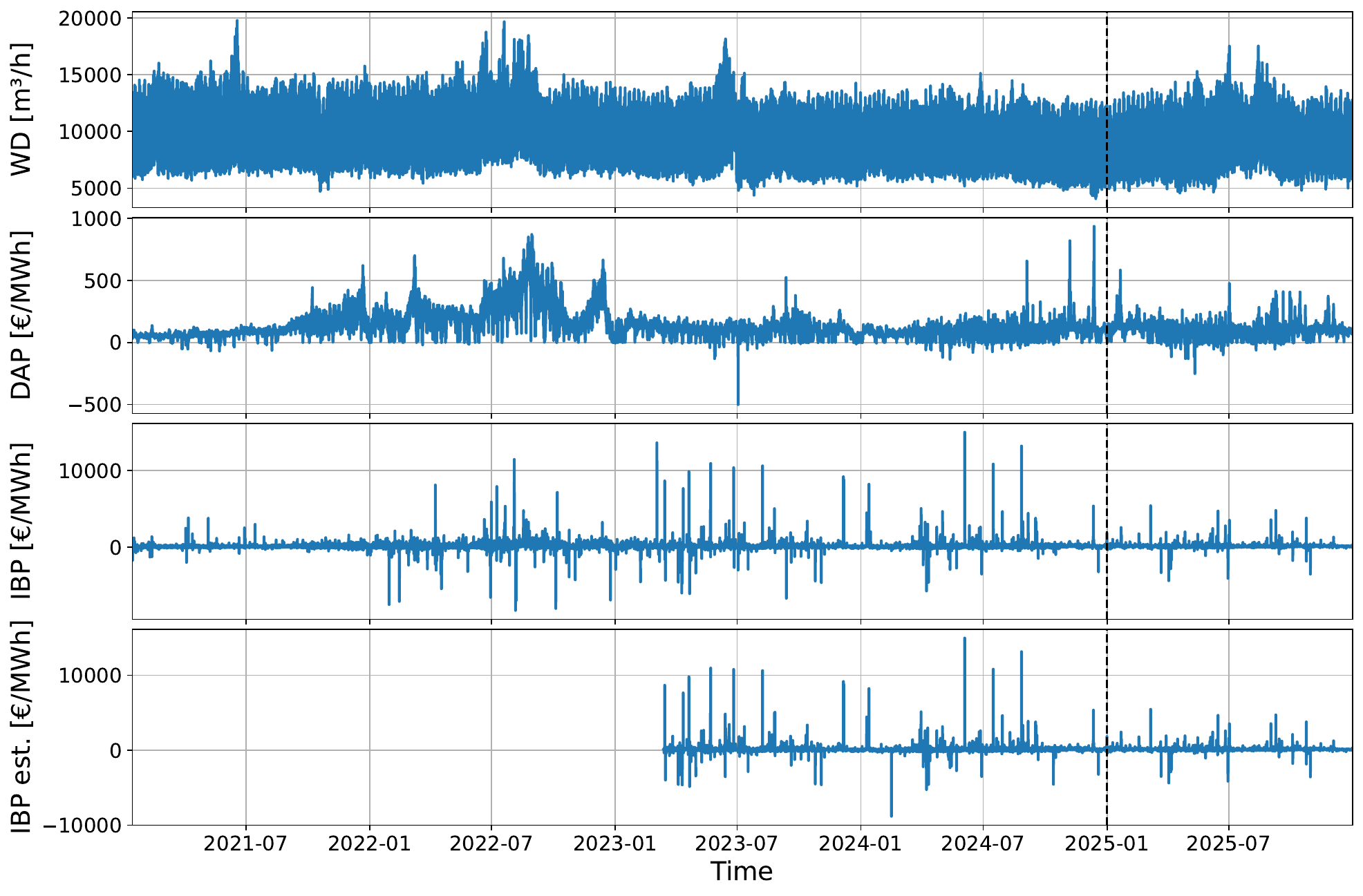}
    \caption{Preprocessed WD, DAP, IBP, and estimate of IBP time series for the training and test periods. Due to limited data availability, IBP estimates are only available from 2023 onwards.}
    \label{fig:inputdata}
\end{figure}

As input data, we employ operational records from a German water supplier; meteorological data provided by the data bases of \textcite{Dwd_mosmix2026} and \textcite{Zippenfenig2023}; as well as energy market data obtained from \textcite{Netztransparenz2026} and \textcite{Entsoe2026}. The data sets cover the period from 2021 to 2025, where the last year is reserved as test data. The operational data of the water supplier includes pump flow, pump head and electricity consumption of the pumps, water volumes of the storage tank, as well as the aggregated water demand. The meteorological data comprise historical and forecast values for temperature, precipitation, cloudiness, and sunshine duration. The energy market data includes DAP as well as IBP and their estimates, load forecasts, and forecasts for RES (wind and solar). A methodological inconsistency arises because RES forecasts are published only after DA gate closure, so their inclusion violates non-anticipativity. Nevertheless, we accept this, as RES forecasts are highly influential for price formation and are widely used in practice, typically via paid data services. All data were resampled to hourly resolution, adjusted for clock changes, and preprocessed using the R package \texttt{tsrobprep} \parencite{Narajewski2021}. The preprocessed input data for the key exogenous processes are shown in Figure \ref{fig:inputdata}.

\section{Methodology}
To represent the sequential stochastic optimization problem, we adopt the unified modeling framework proposed by \textcite{Powell2019}. This framework provides a canonical representation of the stochastic decision problem and supports the systematic benchmarking of control strategies, as advocated by \textcite{Castelletti2023}. The methodology is introduced in four steps. First, the base model is defined, describing the underlying decision problem and system dynamics in their most general form. Second, we present the probabilistic forecasting models used to model the exogenous information processes. Third, we introduce a set of policy models. Finally, we describe the evaluation framework. Throughout the paper, uppercase symbols generally denote stochastic processes, random quantities, or abstract components of the sequential decision framework, whereas lowercase symbols denote corresponding realizations, observations, or deterministic quantities. Forecasted and estimated quantities are indicated using the hat notation.

\subsection{The base model}
\label{sec:basemodel}
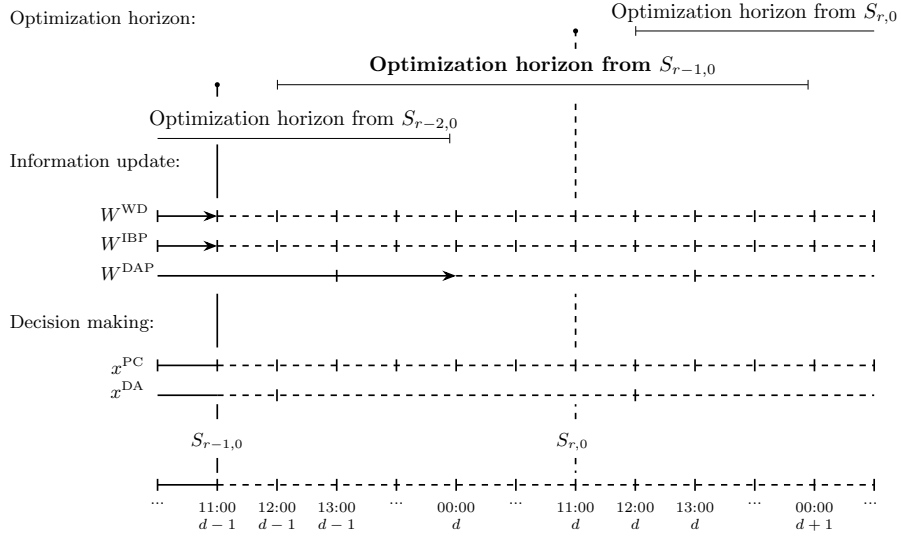
\begin{figure}[ht]
\resizebox{\textwidth}{!}{
\begin{tikzpicture}[
    >=Stealth,
    timeline/.style={thick},
    event/.style={rectangle, align=center, font=\small},
    label/.style={font=\scriptsize}
]


\begin{scope}[yshift=+4.3cm]
\node[event, anchor=west] at (-1.6,3.5)(pricepub) {Optimization horizon:};

\draw[-|] (1, 1.5) -- (5.9, 1.5) node[midway,above]{Optimization horizon from $S_{r-2,0}$};

\draw[|-|] (3,2.4) -- (11.9,2.4) node[midway,above]{\bfseries{Optimization horizon from $S_{r-1,0}$}};

\draw[|-] (9,3.3) -- (13,3.3) node[midway,above]{Optimization horizon from $S_{r,0}$};
\end{scope}

\begin{scope}[yshift=-3.5cm]
\node[event, anchor=west] at (-1.6 ,8.9) (pricepub) {Information update:};

%
%

\draw[timeline][->] (1,8) -- (2,8);
\draw[timeline, dashed] (2,8) -- (13,8);

\tikzset{label/.style={font=\scriptsize, align=center}}

\foreach \x/\t in {
    1/{}, 
    2/{},
    3/{},
    4/{},
    5/{},
    6/{},
    7/{},
    8/{},
    9/{}, 
    10/{},
    11/{}, 
    12/{}, 
    13/{}
} {
    \draw[thick](\x,8.1) -- (\x,7.9);
}

\node[event,above=8 cm of {(0.45,-0.2)}] (pricepub) {$W^{\mathrm{WD}}$};

%
%

\draw[timeline][->] (1,7.5) -- (2,7.5);
\draw[timeline, dashed] (2,7.5) -- (13,7.5);

\tikzset{label/.style={font=\scriptsize, align=center}}

\foreach \x/\t in {
    1/{}, 
    2/{},
    3/{},
    4/{},
    5/{},
    6/{},
    7/{},
    8/{},
    9/{}, 
    10/{},
    11/{}, 
    12/{}, 
    13/{}
} {
    \draw[thick](\x,7.4) -- (\x,7.6);
}
\node[event,above=7.5cm of {(0.45,-0.2)}] (pricepub) {$W^{\mathrm{IBP}}$};

%
%

\draw[timeline][->] (1,7) -- (6,7);
\draw[timeline, dashed] (6,7) -- (13,7);

\foreach \x in {4,10} {
    \draw[thick](\x,6.9) -- (\x,7.1);
}

\node[event, anchor=7.5cm] at (0.1,6.8) {$W^{\mathrm{DAP}}$};
\end{scope}


\begin{scope}[yshift=-3.8cm]
\node[event, anchor=west] at (-1.6,6.5) {Decision making:};

%
%

\draw[timeline] (1,5.8) -- (2,5.8);
\draw[timeline, dashed] (2,5.8) -- (13,5.8);

\foreach \x/\t in {
    1/{}, 
    2/{},
    3/{},
    4/{},
    5/{},
    6/{},
    7/{},
    8/{},
    9/{}, 
    10/{},
    11/{}, 
    12/{}, 
    13/{}
} {
    \draw[thick](\x,5.9) -- (\x,5.7);
}

\node[event] at (0.5,5.8) {$x^{\mathrm{PC}}$};

%
%

\draw[timeline] (1,5.3) -- (2,5.3);
\draw[timeline, dashed] (2,5.3) -- (13,5.3);

\foreach \x in {3,9} {
    \draw[thick](\x,5.2) -- (\x,5.4);
}

\node[event] at (0.5,5.4) {$x^{\mathrm{DA}}$};
\end{scope}


\draw[thick] (2,1.1) -- (2,1.35);
\draw[thick] (2,2.3) -- (2,3.2);
\draw[thick] (2,4.8) -- (2,5.7);
\draw[thick] (2,6.5) -- (2,6.7);
\fill (2,6.7) circle (1.2pt);

\draw[thick, dashed] (8,1.1) -- (8,1.35);
\draw[thick, dashed] (8,2.3) -- (8,3.2);
\draw[thick, dashed] (8,4.8) -- (8,6.4);
\draw[thick, dashed] (8,7.3) -- (8,7.6);
\fill (8,7.6) circle (1.2pt);


\draw[timeline] (1,0) -- (2,0);
\draw[timeline, dashed] (2,0) -- (13,0);;
\tikzset{label/.style={font=\scriptsize, align=center}}

\foreach \x/\t in {
    1/{...}, 
    2/{11:00\\$d-1$},
    3/{12:00\\$d-1$},
    4/{13:00\\$d-1$},
    5/{...},
    6/{00:00\\$d$},
    7/{...},
    8/{11:00\\$d$},
    9/{12:00\\$d$}, 
    10/{13:00\\$d$},
    11/{...}, 
    12/{00:00\\$d+1$}, 
    13/{...}
} {
    \draw[thick] (\x,0.1) -- (\x,-0.1);
    \node[label,below=2pt] at (\x,-0.1) {\t};
}

\node[event,above=0.15cm of {(2,0.3)}] (pricepub) {$S_{r-1,0}$};
\draw[-, thick] (pricepub.south) -- (2,0.2);

\node[event,above=0.15cm of {(8,0.3)}] (pricepub) {$S_{r,0}$ };
\draw[dashed,thick] (pricepub.south) -- (8,0.2);

\end{tikzpicture}
} 

\caption{Receding-horizon structure of consecutive optimization runs and their temporal overlap, together with the sequence of information revelation and decision making for the water demand process $W^{\mathrm{WD}}$, the imbalance price process $W^{\mathrm{IBP}}$, and the day-ahead price process $W^{\mathrm{DAP}}$.}
\label{fig:sequence of events}
\end{figure}

The base model is aligned to the sequence of events illustrated in Figure \ref{fig:sequence of events}, which depicts the timing of optimization, information revelation, and decision making. For simplicity, the optimization run index $r$ is omitted in the following, and only the index $t$, representing the internal time step within a single optimization run, is retained.

The sequential decision problem is described by the state variables $S_t$, the decision variables $x_t$, the exogenous information process $W_{t+1}$, the transition function $S^M$, and the objective function. Rather than optimizing directly over decisions \(x_t\), the framework optimizes over policies \(\pi\), where decisions are determined by
\begin{equation}
x_t = X_t^\pi(S_t),
\end{equation}
where $X_t(\cdot)$ denotes the decision function under policy $\pi$, mapping the state $S_t$ to the decision $x_t$.

\subsubsection{State variable}
The state variables include everything the decision depends on and all information necessary to let the system evolve via the transition and contribution functions. The state variable can be defined as:

\begin{equation}
S_t = (V_{t}, x_{t-1})
\end{equation}

where $V_t$ denotes the storage volume and $x_{t-1}$ the previous decision vector. We distinguish between the state $S_t$ for $t>0$ and the initial state $S_0$. The latter contains both the initial values of the dynamic state variables and all deterministic parameters required to define the system. It is given by
\begin{equation}
S_0 = (V_0, ...).
\end{equation}

\subsubsection{Decision variables}
The decision at time $t$ is represented by a vector
\begin{equation}
x_t = \left(x_t^{\mathrm{DA}},\, x_t^{\mathrm{PC}}\right),
\end{equation}
where $x_t^{\mathrm{DA}}$ denotes the day-ahead procurement decision and $x_t^{\mathrm{PC}}$ the pump operation decision. At each time step, exactly one pump combination must be selected, i.e.,
\begin{equation}
x_t^{\mathrm{PC}} \in \mathcal{X},
\end{equation}
where $\mathcal{X}$ denotes the set of all available pump combinations for the corresponding optimization run. The two components of the decision vector are coupled through the resulting electricity consumption. While the day-ahead procurement decision $x_t^{\mathrm{DA}}$ must be fixed prior to the realization of operational uncertainty, the pump operation decision $x_t^{\mathrm{PC}}$ can be updated sequentially as uncertainty unfolds and additional information becomes available.

The deviation between realized electrical consumption and procured electricity is given by
\begin{equation}
I_t = P(S_t, x_t^{\mathrm{PC}}) - x_t^{\mathrm{DA}},
\end{equation}
where the electricity consumption $P(S_t, x_t^{\mathrm{PC}})$ depends only on the
operational component $x_t^{\mathrm{PC}}$. To exclude systematic arbitrage between the day-ahead and imbalance market, these deviations are assumed to have zero mean under the policy $\pi$, i.e.,
\begin{equation}
\mathbb{E}^{\pi}[I_t] = 0, \qquad \forall t.
\label{eq:zeroimbalance}
\end{equation}
Consequently, imbalance costs arise solely from uncertainty-driven deviations rather than deliberate market positioning.

\subsubsection{Exogenous information}
The exogenous information process $W_{t+1}$ represents the stochastic uncertainty associated with decision interval $t$ after implementing decision $x_t$ and before reaching state $S_{t+1}$. Consequently, the quantities contained in $W_{t+1}$ affect both the transition from $S_t$ to $S_{t+1}$ and the realized contribution of decision $x_t$ to the objective function. It is modeled as the vector-valued random variable
\begin{equation}
W_{t+1} = \left(W_{t+1}^{\mathrm{WD}},\, W_{t+1}^{\mathrm{DAP}},\, W_{t+1}^{\mathrm{IBP}}\right),
\end{equation}
where $W_{t+1}^{\mathrm{WD}}$ denotes the stochastic water demand process, $W_{t+1}^{\mathrm{DAP}}$ the stochastic day-ahead electricity price process, and $W_{t+1}^{\mathrm{IBP}}$ the stochastic imbalance price process associated with decision interval $t$. Realizations of the exogenous processes are denoted by the corresponding lowercase quantities $w_{t+1}^{\mathrm{WD}}$, $w_{t+1}^{\mathrm{DAP}}$, and $w_{t+1}^{\mathrm{IBP}}$. The individual exogenous processes differ with respect to the timing of their realization, publication, and operational relevance. Future realizations of the exogenous information process are unknown at decision time. Consequently, policies operating under uncertainty require forecast representations of future realizations of $W_{t+1}$. The exogenous processes are assumed to be independent of the system state and the control decisions. In particular, the water supplier is treated as a price taker such that operational decisions do not affect price formation in either the day-ahead or imbalance market. Likewise, water demand is assumed to be independent of supply decisions.

\subsubsection{Transition function}
The transition function describes the evolution of the system state according to
\begin{equation}
S_{t+1} = S^M(S_t, x^{\mathrm{PC}}_t, W_{t+1}).
\end{equation}

Specifically, the storage volume is represented by a mass balance
\begin{equation}
V_{t+1} = V_t + Q(S_t,x^{\mathrm{PC}}_t) - W^{\mathrm{WD}}_{t+1},
\end{equation}
where \(Q(S_t,x^{\mathrm{PC}}_t)\) denotes the flow rate associated with the selected pump combination \(x^{\mathrm{PC}}_t\).

Following \textcite{Menke2016} and \textcite{Bonvin2021}, pump flow and electricity consumption are approximated by affine relationships. The flow rate is defined as
\begin{equation}
Q(S_t, x^{\mathrm{PC}}_t)
=
\beta^{Q}_{0,x^{\mathrm{PC}}_t}
+
\beta^{Q}_{1,x^{\mathrm{PC}}_t} H_t,
\qquad
\forall x^{\mathrm{PC}}_t \in \mathcal{X},
\end{equation}
while the corresponding electricity consumption is given by
\begin{equation}
P(S_t, x^{\mathrm{PC}}_t)
=
\beta^{P}_{0,x^{\mathrm{PC}}_t}
+
\beta^{P}_{1,x^{\mathrm{PC}}_t} Q(S_t, x^{\mathrm{PC}}_t),
\qquad
\forall x^{\mathrm{PC}}_t \in \mathcal{X}.
\end{equation}

System head is modeled as an affine function of storage volume:
\begin{equation}
H_t
=
h^{\mathrm{min}}
+
\frac{(V_t - v^{\mathrm{min}})(h^{\mathrm{max}} - h^{\mathrm{min}})}
{v^{\mathrm{max}} - v^{\mathrm{min}}},
\end{equation}
where \(v^{\mathrm{min}}\) and \(v^{\mathrm{max}}\) denote the physical storage capacities. The parameters \(h^{\mathrm{min}}\) and \(h^{\mathrm{max}}\) are derived from historical observations and correspond to the \(0.05\) and \(0.95\) quantiles of observed head values.

To ensure reliable operation of the storage system and pump combinations, the operating ranges are constructed in accordance to Figure~\ref{fig:operating_ranges}. The bounds are assumed to be either operator-defined or derived from data.

For the storage system, the upper bounds \(V_t^{\mathrm{uba}}\) and \(V_t^{\mathrm{ubc}}\) are specified as fixed operator-defined thresholds preventing overflow. In contrast, the lower bounds are determined in a risk-aware manner based on the distribution of future water demand under the assumption of zero inflow, e.g., caused by pump failure. Specifically,
\begin{equation}
V^{\kappa}_{t}
=
\mathrm{CVaR}_{\alpha^{\kappa}}
\!\left(
\sum_{i=0}^{b^{\kappa}} W^{\mathrm{WD}}_{t+i}
\right),
\qquad
\kappa \in \{\mathrm{lba}, \mathrm{lbc}\},
\label{eq:storageboundscvar}
\end{equation}
where \(b^{\kappa}\) denotes the operator-defined intervention horizon in hours associated with the respective threshold. In the present work,
\(b^{\mathrm{lba}} = 1\) and \(b^{\mathrm{lbc}} = 0.75\), while
\(\alpha^{\mathrm{lba}} = \alpha^{\mathrm{lbc}} = 0.95\).
The tail-risk measure is formally introduced in the objective-function section.

The feasible operating ranges of each pump combination are derived from historical operating data. For each pump combination, \(h^{\mathrm{lba}}_{PC}\) and \(h^{\mathrm{uba}}_{PC}\) are defined as the \(0.05\) and \(0.95\) empirical quantiles of the observed head values, whereas \(h^{\mathrm{lbc}}_{PC}\) and \(h^{\mathrm{ubc}}_{PC}\) correspond to the minimum and maximum observed values. The quantities \(h^{\mathrm{lbc}}_{PC}\), \(h^{\mathrm{ubc}}_{PC}\), \(h^{\mathrm{lba}}_{PC}\), and \(h^{\mathrm{uba}}_{PC}\) therefore denote deterministic operating thresholds. In the objective function, these quantities are accessed through deterministic mappings, which return the corresponding threshold associated with the selected pump combination \(x_t^{\mathrm{PC}}\).

\subsection{Objective function}
The objective is formulated as a search over policies $\pi \in \Pi$ and minimizes total operational costs over the planning horizon. It accounts for economic costs, smooth pump operation, and system reliability, while incorporating risk aversion with respect to economic outcomes. The objective function is defined as:
\begin{equation}
\begin{aligned}
\min_{\pi \in \Pi} \quad
&(1-\lambda^{\mathrm{risk}})\,\mathbb{E}^{\pi}\!\left[\sum_{t=0}^{T} C_t^{\mathrm{econ}}(S_t, X_t^{\pi}(S_t), W_{t+1}) | S_0\right] \\
&+ \lambda^{\mathrm{risk}}\,\mathrm{CVaR}_{\alpha^{\mathrm{econ}}}^{\pi}\!\left(\sum_{t=0}^{T} C_t^{\mathrm{econ}}(S_t, X_t^{\pi}(S_t), W_{t+1})| S_0\right) \\
&+ \mathbb{E}^{\pi}\!\left[\sum_{t=0}^{T} C_t^{\mathrm{smo}}(S_t, X_t^{\pi}(S_t))| S_0\right] \\
&+ \sum_{t=0}^{T} C_t^{\mathrm{rel}}(S_t, X_t^{\pi}(S_t)),
\end{aligned}
\label{eq:objective}
\end{equation}
where $\lambda^{\mathrm{risk}} \in [0,1]$ controls the trade-off between expected cost and economic risk. The reliability term is evaluated through scenario-based exceedance-risk measures (ER) at each time step and is therefore already aggregated across uncertainty. Consequently, it enters the objective as a stage-wise risk penalty.

To capture risk-averse behavior, risk measures based on the contribution of \textcite{Rockafellar2000} are employed. 
For economic performance, risk is modeled using the standard Conditional Value-at-Risk (CVaR) formulation:
\begin{equation}
\mathrm{CVaR}_{\alpha}(Z)
=
\min_{\zeta \in \mathbb{R}}
\left\{
\zeta
+
\frac{1}{1-\alpha}
\mathbb{E}\!\left[(Z - \zeta)^+\right]
\right\},
\end{equation}
where $Z$ denotes a real-valued loss variable, $\zeta$ is an auxiliary variable corresponding to the Value-at-Risk (VaR) at confidence level $\alpha$, and $(u)^+ := \max(0,u)$ denotes the positive part of a real-valued quantity. Since $Z$ represents costs, the CVaR term captures the upper tail of the cost distribution and thus penalizes scenarios with particularly high operational costs.

The CVaR formulation is employed in two conceptually distinct ways within the proposed framework. First, it is used exogenously in Equation~\ref{eq:storageboundscvar} within the transition-function section to derive risk-aware storage thresholds from forecast demand distributions. In this case, the resulting quantities constitute fixed parameters of the system model. Second, CVaR is incorporated directly into the objective function to model risk aversion with respect to economic costs, where the associated auxiliary variable is optimized jointly with the policy.

For reliability-related quantities, admissible operating-limit violations are penalized using normalized violation measures. For a normalized violation measure $\tilde{Z}$, this is defined as
\begin{equation}
\mathrm{ER}_{\alpha}(\tilde{Z})
=
\frac{1}{1-\alpha}
\mathbb{E}\!\left[(\tilde{Z})^+\right].
\end{equation}
In contrast to the CVaR formulation, which determines a risk threshold endogenously through the optimization variable $\zeta$, the ER formulation evaluates exceedances relative to exogenously specified operating limits. Consequently, ER does not optimize a threshold but measures the expected magnitude of violations beyond a fixed admissible limit.

The economic contribution accounts for DA procurement and imbalance costs:
\begin{equation}
C_t^{\mathrm{econ}}(S_t,x_t,W_{t+1}) =  W^{\mathrm{DAP}}_{t+1} \cdot x_t^{\mathrm{DA}} +
 W^{\mathrm{IBP}}_{t+1} \cdot I_t,
\end{equation}
where $I_t$ denotes the deviation between realized electrical consumption $P(S_t,x^{\mathrm{PC}}_t)$ and the DA commitment $x^{\mathrm{DA}}_t$.  
Due to the market-neutrality condition $I_t$ has zero mean, implying that imbalance costs arise solely from the joint realization of demand and imbalance prices rather than deliberate positioning.
The pump start-up contribution penalizes the activation of pumps to promote stable and smooth pump operation:
\begin{equation}
C_t^{\mathrm{smo}}(S_t, x^{\mathrm{PC}}_t) =
\sum_{p \in \mathcal{P}} c^{\mathrm{smo}} \cdot 
\mathbf{1}\!\left\{ p \notin x_{t-1}^{\mathrm{PC}},\, p \in x_t^{\mathrm{PC}} \right\},
\end{equation}
where $\mathbf{1}\{\cdot\}$ denotes the indicator function, which equals $1$ if the corresponding pump is selected and $0$ otherwise.

The system reliability contribution penalizes expected exceedances of admissible operating ranges based on normalized violations for the storage volume and the pump-combination-specific head. We assume a symmetric structure for upper and lower operating ranges for storage and pump head, respectively:
\begin{equation}
C_t^{\mathrm{rel}}(S_t, x^{\mathrm{PC}}_t) =
c_t^{\mathrm{cor}} \sum_{k \in \mathcal{K}} 
\mathrm{ER}_{\alpha^{\mathrm{rel}}}(\tilde{Z}_t^k)
+
c^{\mathrm{eor}} \sum_{k \in \mathcal{K}}
\left(\mathrm{ER}_{\alpha^{\mathrm{rel}}}(\tilde{Z}_t^k) - 1\right)^+,
\label{eq:reliabilitycontribution}
\end{equation}
where
\[
\mathcal{K} = \{\mathcal{V}_{\mathrm{low}},\, \mathcal{V}_{\mathrm{up}},\, 
\mathcal{H}_{\mathrm{low}},\, \mathcal{H}_{\mathrm{up}}\}.
\]
Here, $\tilde{Z} \leq 0$ corresponds to operation within the admissible and $0 \leq \tilde{Z} \leq 1$ to operation within critical operating range, while values exceeding $1$ indicate severe violations within the emergency operating domain.
The system reliability penalty works in two regimes. For moderate violations within the critical operating range, violations are penalized linearly with weight $c^{\mathrm{cor}} = 30$, allowing such deviations only when they are economically justified or necessary to avoid infeasibility. The second regime penalizes severe violations beyond the critical operating range with a substantially larger weight, thereby enforcing feasibility-driven control decisions. $c^{\mathrm{eor}}$ is derived by estimating the upper bound along the optimization horizon for joint economic and switching costs.
 
To ensure trajectory consistency, violations are evaluated over transitions, i.e., over the set $\{t-1,t\}$.

For the storage volume, they are defined as:
\begin{equation}
\tilde{Z}_t^{\mathcal{V}_{\mathrm{low}}}
=
\frac{\max_{\tau \in \{t-1,t\}} \left( V_t^{\mathrm{lba}} - V_{\tau} \right)^+}{\varepsilon^{\mathcal{V}_{\mathrm{low}}}(S_t)},
\end{equation}
\begin{equation}
\tilde{Z}_t^{\mathcal{V}_{\mathrm{up}}}
=
\frac{\max_{\tau \in \{t-1,t\}} \left( V_{\tau} - V_t^{\mathrm{\text{ubc}}} \right)^+}{\varepsilon^{\mathcal{V}_{\mathrm{up}}}(S_t)},
\end{equation}
where $\varepsilon^{\mathcal{V}_{\mathrm{low}}}$ and $\varepsilon^{\mathcal{V}_{\mathrm{up}}}$ denote the critical operating ranges:
\begin{equation}
\varepsilon^{\mathcal{V}_{\mathrm{low}}}(S_t) = V_t^{\mathrm{lba}} - V_t^{\mathrm{lbc}}, 
\qquad
\varepsilon^{\mathcal{V}_{\mathrm{up}}}(S_t)  = v_t^{\mathrm{ubc}} - v_t^{\mathrm{\text{uba}}}.
\end{equation}
The lower bounds $V^{\mathrm{lbc}}$ and $V^{\mathrm{lba}}$ are derived by applying Equation \ref{eq:storageboundscvar}. The normalized violation variables  pump head are defined analogously:
\begin{equation}
\tilde{Z}_t^{\mathcal{H}_{\mathrm{low}}}
=
\frac{\max_{\tau \in \{t-1,t\}} \left( H^{\mathrm{lb}}(x_t^{\mathrm{PC}}) - H_{\tau} \right)^+}{\varepsilon^{\mathcal{H}_{\mathrm{low}}}(x_t^{\mathrm{PC}})},
\end{equation}
\begin{equation}
\tilde{Z}_t^{\mathcal{H}_{\mathrm{up}}}
=
\frac{\max_{\tau \in \{t-1,t\}} \left( H_{\tau} - H^{\mathrm{\text{ubc}}}(x_t^{\mathrm{PC}}) \right)^+}{\varepsilon^{\mathcal{H}_{\mathrm{up}}}(x_t^{\mathrm{PC}})}.
\end{equation}

where $\varepsilon^{\mathcal{H}_{\mathrm{low}}}$ and $\varepsilon^{\mathcal{H}_{\mathrm{up}}}$ denote the critical operating ranges:
\begin{equation}
\begin{aligned}
\varepsilon^{\mathcal{H}_{\mathrm{low}}}(x_t^{\mathrm{PC}}) &=
H^{\mathrm{lba}}(x_t^{\mathrm{PC}}) - H^{\mathrm{lbc}}(x_t^{\mathrm{PC}}), \\
\varepsilon^{\mathcal{H}_{\mathrm{up}}}(x_t^{\mathrm{PC}})  &=
H^{\mathrm{ubc}}(x_t^{\mathrm{PC}}) - H^{\mathrm{\text{uba}}}(x_t^{\mathrm{PC}}),
\end{aligned}
\end{equation}
The functions \(H^{\mathrm{lbc}}(x_t^{\mathrm{PC}})\), \(H^{\mathrm{lba}}(x_t^{\mathrm{PC}})\), \(H^{\mathrm{uba}}(x_t^{\mathrm{PC}})\), and \(H^{\mathrm{ubc}}(x_t^{\mathrm{PC}})\) map the selected pump combination \(x_t^{\mathrm{PC}}\) to the corresponding operating thresholds \(h^{\mathrm{lbc}}_{PC}\), \(h^{\mathrm{lba}}_{PC}\), \(h^{\mathrm{uba}}_{PC}\), and \(h^{\mathrm{ubc}}_{PC}\).

\begin{table}[ht]
\centering
\caption{Principal notation of the base model of Section \ref{sec:basemodel}}
\label{tab:base_model_notation}

\begin{tabular}{p{0.22\linewidth} p{0.58\linewidth} p{0.12\linewidth}}
\hline
Symbol & Description & Unit \\
\hline

\multicolumn{3}{l}{\textbf{Indices and sets}} \\
$t$ & Time step index & h \\
$r$ & Optimization run index & - \\
$\Pi$ & Set of admissible policies & - \\
$\mathcal{X}$ & Set of feasible PCs & - \\
$\mathcal{P}$ & Set of pumps & - \\
$\mathcal{K}$ & Set of reliability constraints & - \\
\hline

\multicolumn{3}{l}{\textbf{Endogenous variables}} \\
$V_t$ & Storage volume & $\mathrm{m}^3$ \\
$H_t$ & System head & m \\
$x_t^{\mathrm{DA}}$ & Day-ahead procurement decision & MWh \\
$x_t^{\mathrm{PC}}$ & Pump-combination decision & - \\
$I_t$ & Imbalance energy & MWh \\
\hline

\multicolumn{3}{l}{\textbf{Exogenous information}} \\
$W_t^{\mathrm{WD}}$ & Water demand process & $\mathrm{m}^3/\mathrm{h}$ \\
$W_t^{\mathrm{DAP}}$ & Day-ahead electricity price process & EUR/MWh \\
$W_t^{\mathrm{IBP}}$ & Imbalance price process & EUR/MWh \\
$V_t^{\mathrm{lba}},V_t^{\mathrm{lbc}}$ & Forecast-derived lower storage bounds & $\mathrm{m}^3$ \\
$c^{\mathrm{eor}}$ & Forecast-derived emergency operating cost & EUR \\
\hline

\multicolumn{3}{l}{\textbf{Functions}} \\
$Q(\cdot)$ & Flow rate function per PC & $\mathrm{m}^3/\mathrm{h}$ \\
$P(\cdot)$ & Electricity consumption function per PC & MWh \\
$C^{\mathrm{econ}}(\cdot)$ & Economic contribution function & EUR \\
$C^{\mathrm{smo}}(\cdot)$ & Smooth-operation contribution function & EUR \\
$C^{\mathrm{rel}}(\cdot)$ & Reliability contribution function & EUR \\
\hline

\multicolumn{3}{l}{\textbf{Parameters}} \\
$\lambda^{\mathrm{risk}}$ & Economic risk-aversion parameter & - \\
$\alpha^{\mathrm{econ}},\alpha^{\mathrm{rel}}$ & Economic and reliability confidence levels & - \\
$\alpha^{\mathrm{lbc}},\alpha^{\mathrm{lba}}$ & Lower storage bounds confidence levels & - \\
$c^{\mathrm{psc}}$ & Pump start-up cost & EUR \\
$c^{\mathrm{cor}}$ & Critical operating range penalty cost & EUR \\
$v^{\mathrm{min}},v^{\mathrm{max}}$ & Minimum and maximum storage capacity & $\mathrm{m}^3$ \\
$h^{\mathrm{min}},h^{\mathrm{max}}$ & Head corresponding to $v^{\mathrm{min}}$ and $v^{\mathrm{max}}$ & m \\
$v_t^{\mathrm{uba}},v_t^{\mathrm{ubc}}$ & Upper storage bounds & $\mathrm{m}^3$ \\
$h_{PC}^{\mathrm{lba}},h_{PC}^{\mathrm{lbc}}$ & Lower head bounds per PC & m \\
$h_{PC}^{\mathrm{uba}},h_{PC}^{\mathrm{ubc}}$ & Upper head bounds per PC & m \\
$b^{\mathrm{lba}},b^{\mathrm{lbc}}$ & Intervention horizons & h \\
\hline

\end{tabular}
\end{table}

\begin{table}[ht]
\centering
\caption{Parameterization of the numerical study.}
\label{tab:numerical_study}

\begin{tabular}{llll}
\hline
Parameter & Description & Unit & Value \\
\hline
\multicolumn{4}{l}{\textbf{Study settings}} \\
$T$ & Optimization horizon & h & 36 \\
$N^{\mathrm{runs}}$ & Optimization runs & - & 100 \\
$N^{\mathrm{reps}}$ & Repetitions & - & 3 \\
\multicolumn{4}{l}{\textbf{Policy approximation settings}} \\
$N^{\mathrm{PC}}$ & Preselected PCs & - & 9 \\
$M^{\mathrm{full}}$ & Full ensemble members & - & 1000 \\
$M^{\mathrm{red}}$ & Reduced scenario members & - & 5 \\
\multicolumn{4}{l}{\textbf{Risk measure settings}} \\
$\lambda^{\mathrm{risk}}$ & Economic risk-aversion parameter & - & 0.5 \\
$\alpha^{\mathrm{econ}}$  & Economic CVaR confidence level & - & 0.8 \\
$\alpha^{\mathrm{rel}}$ & Reliability ER confidence level &  - & 0.8 \\
$\alpha^{\mathrm{lba}}$,$\alpha^{\mathrm{lbc}}$ & $V_t^{\mathrm{lba}}$ and $V_t^{\mathrm{lbc}}$ CVaR confidence level  & - & 0.95, 0.95 \\
\multicolumn{4}{l}{\textbf{Operational control settings}} \\
$b^{\mathrm{lbc}},\,b^{\mathrm{lba}}$ & $V_t^{\mathrm{lba}}$ and $V_t^{\mathrm{lbc}}$ intervention horizons & h & 0.75, 1 \\
$c^{\mathrm{cor}}$ & Critical operating range penalty cost & EUR & 30 \\
$c^{\mathrm{psc}}$ & Pump start-up penalty cost & EUR & 10 \\
$v^{\mathrm{uba}}_t,\,v^{\mathrm{ubc}}_t$ & Upper storage bounds & $\mathrm{m}^3$ & 41000, 44000 \\
\multicolumn{4}{l}{\textbf{Storage system settings}} \\
$v^{\mathrm{min}},\,v^{\mathrm{max}}$ & Storage capacity limits & $\mathrm{m}^3$ & 2000, 44000 \\
$h^{\mathrm{min}},\,h^{\mathrm{max}}$ & Quantile-derived head limits & m & 92.62, 102.40 \\
$V_0$ & Initial storage volume & $\mathrm{m}^3$ & 27000 \\
\hline
\end{tabular}
\end{table}

\subsection{Probabilistic forecast generation}
To model the exogenous information $W_t = (W^{\mathrm{WD}}_t, W^{\mathrm{DAP}}_t, W^{\mathrm{IBP}}_t)$ we employ established approaches from the electricity price and water demand forecasting literature. We consider both simple benchmark models and more advanced methods explicitly designed for probabilistic forecasting. While water demand and day-ahead prices are well-established forecasting targets, imbalance prices are less studied and harder to predict, especially at the day-ahead stage \parencite{Browell2022}. This is due to the fact that much of the relevant information becomes available short to delivery and hence, usually after the DA gate closure. As a result, predictive accuracy is generally low, and even simple benchmarks—such as the DA price—are difficult to outperform. The water demand models are trained on four years of data, whereas the DAP and IBP models are estimated using two years of data, excluding the extreme price period during the 2022 energy crisis to avoid biased estimation. As forecast horizons we consider for the water demand and the imbalance price process 36 hours and for the day-ahead price process 24 hours lead time. The finally obtained forecast ensemble contains $M^{\mathrm{full}}=1000$ trajectories, indexed by \(m=1,\ldots,M^{\mathrm{full}}\). In the proceeding we first derive the marginal forecasting models for each processs $W^{\mathrm{WD}}$, $W^{\mathrm{DAP}}$ and $W^{\mathrm{IBP}}$ and subsequently merge them into one joint distribution $W$.

\subsubsection{Naive models}
The day-ahead electricity price process and the water demand process are modeled as independent processes, applying for the $W^{\mathrm{DAP}}$ the naive benchmark of \textcite{Marcjasz2020} and for the $W^{\mathrm{WD}}$ process the "mean" model, following \textcite{Alvisi2025} and \textcite{Gagliardi2017}. The $W^{\mathrm{IBP}}$ process in contrast is constructed by reusing the point prediction of the naive day-ahead forecasting model and the already published DAP realizations of the current day. Following \textcite{Marcjasz2023}, all point forecasts are converted into probabilistic forecasts via bootstrap resampling of in-sample residual trajectories. These trajectories are generated conditional on the forecasting origin and match the length of the respective forecast horizon. The resulting models are denoted as $\mathrm{DAPF\_Naive}$, $\mathrm{WDF\_Naive}$, and $\mathrm{IBPF\_Naive}$.

\subsubsection{Distributional regression models}
We distinguish between models originally developed for point forecasting and subsequently extended to probabilistic settings, and fully distributional regression models that directly estimate the predictive distribution. In the following, we refer to these as LEAR-based and GAMLSS-based models, respectively.

\paragraph{LEAR-based models.}
The first class is based on the LASSO-estimated autoregressive (LEAR) framework as applied by \textcite{Marcjasz2023} and \textcite{Lago2021}. Here, the least absolute shrinkage and selection operator (LASSO) \parencite{Tibshirani1996} is used to regularize a high-dimensional feature space and induce sparsity. Estimation is performed via the least angle regression (LARS) algorithm \parencite{Efron2004}, where the regularization parameter $\lambda^{\text{LASSO}}$ is selected using the Bayesian Information Criterion (BIC), following \textcite{Lago2021}. The models are estimated using the implementation provided by \texttt{scikit-learn} \parencite{Pedregosa2011}.

\paragraph{GAMLSS-based models.}
The second class follows the generalized additive model for location, scale, and shape (GAMLSS) framework \parencite{Rigby2005}, which directly models the full predictive distribution. Let $Y_i$, $i = 1, \ldots, n$, denote independent observations of the response variable. The model is specified as
\begin{equation}
Y_i \sim \mathcal{D}(\theta_{i,1}, \ldots, \theta_{i,p}),
\end{equation}
where $\mathcal{D}$ denotes a parametric distribution with up to $p$ parameters. Each parameter $\theta_{i,k}$ is linked to a predictor via
\begin{equation}
g_k(\theta_{i,k}) = \eta_{i,k} = \mathbf{x}_{i,k}^\top \boldsymbol{\beta}_k, \quad k = 1, \ldots, p,
\end{equation}
allowing each distributional parameter to depend on its own set of covariates. This enables flexible modeling of location, scale, and higher-order shape characteristics.
Estimation is performed via (penalized) maximum likelihood using the Rigby–Stasinopoulos algorithm. To address high-dimensional feature spaces, LASSO regularization can be applied within parameter-specific regressions \parencite{Groll2019, Ziel2022}. For implementation, we use the Python package \texttt{ondil} \parencite{Hirsch2024}, which supports distributional regression with regularization. In this work, models are estimated in a batch setting using LASSO-based feature selection with BIC tuning. As distributional assumptions, we consider the normal and Student’s t distribution. The corresponding parameters are denoted by $\mu$ (location), $\sigma$ (scale), and $\nu$ (shape). The identity link is used for $\mu$, while a logarithmic link is applied for $\sigma$. For the Student’s t distribution, we specify $g_{\nu}(\nu) = \log(\nu - 2)$ to ensure $\nu > 2$. Compared to the normal distribution allows the Student’s t for greater flexibility in capturing heavy-tailed behavior.

\subsubsection{Day-ahead price forecasting models}
Given that day-ahead prices for all hourly products are determined simultaneously, we estimate separate models for each hour of the day. The resulting marginal predictive distributions are subsequently combined by modeling the temporal dependence structure via a Gaussian copula to approximate the process-specific multivariate forecast distribution. Finally, the required trajectories are sampled from the resulting distribution.

\paragraph{LEAR-based model.}
Following \textcite{Marcjasz2023}, we estimate four independent LEAR models and subsequently apply Quantile Regression Averaging (QRA) to obtain marginal forecast distributions for each hour. The LEAR models differ in their degree of temporal adaptivity through rolling training windows of 457, 548, 639, and 730 days, respectively. For each training window we define one model for each hour $h$ of the day:
\begin{equation}
\begin{aligned}
\hat{\mu}_{d,h} =\;& \beta_{\mu,0,h} 
+ \sum_{h'=1}^{24} \beta_{\mu,h',h}\, w^{\mathrm{DAP}}_{d-1,h'}
+ \sum_{h'=1}^{24} \beta_{\mu,24+h',h}\, w^{\mathrm{DAP}}_{d-2,h'} \\
&+ \sum_{h'=1}^{24} \beta_{\mu,48+h',h}\, w^{\mathrm{DAP}}_{d-3,h'}
+ \sum_{h'=1}^{24} \beta_{\mu,72+h',h}\, w^{\mathrm{DAP}}_{d-7,h'} \\
&+ \sum_{i=1}^{7} \beta_{\mu,97+i,h}\, DoW_{d,i}
+ \beta_{\mu,105,h}\, \widehat{\text{Load}}_{d,h} \\
&+ \sum_{h'=1}^{24} \beta_{\mu,105+h',h}\, \widehat{\text{RES}}_{d,h'} ,
\end{aligned}
\label{eq:dapf_location}
\end{equation}
where $\hat{\mu}_{d,h}$ denotes the point forecast (location parameter) of the day-ahead price for hour $h$ on day $d$, and $\beta_{\mu,i,h}$ are the corresponding coefficient parameters. The variables $DoW_{d,i}$ represent day-of-week dummies, while $\widehat{\text{RES}}$ and $\widehat{\text{Load}}$ denote forecasts of renewable generation and load, respectively. The variable $w^{\mathrm{DAP}}$ represents a vector with historical DAP realizations. The marginal probabilistic forecasts for each hour $h$ are obtained by applying QRA with a calibration window of 90 days, where the four independent point forecasts used as covariates. We refer to this model as $\mathrm{DAPF\_LEAR\_QRA}$.
\\

\paragraph{GAMLSS-based models.}
We consider specifications based on the Student's t distribution to account for the heavy tailed price data. The location predictor $\eta_{\mu,h}$ is specified in accordance with Equation~\eqref{eq:dapf_location}. The predictors for the remaining distributional parameters $\eta_{k,h}$, $k \neq \mu$, are given by
\begin{equation}
\begin{aligned}
\eta_{k,h} =\;& \beta_{k,0,h} 
+ \beta_{k,1,h}\, \bar{w}^{\mathrm{DAP}}_{d-1}
+ \sum_{i=1}^{7} \beta_{k,1+i,h}\, DoW_{d,i} \\
&+ \beta_{k,9,h}\, \text{MAD}^{[d-7,d-1]}(w^{\mathrm{DAP}})
+ \beta_{k,10,h}\, \widehat{\text{Load}}_{h}\\
&+ \beta_{k,11,h}\, \widehat{\text{RES}}_{h},
\end{aligned}
\label{eq:dapf_higher_moments}
\end{equation}
where $\bar{w}^{\mathrm{DAP}}_{d-1}$ denotes the mean of the previous day DAP vector, and $\text{MAD}^{[d-7,d-1]}(w^{\mathrm{DAP}})$ the mean absolute deviation computed over the past seven days. The index $k$ refers to the respective distributional parameter. The resulting model are denoted as $\mathrm{DAPF\_GAMLSS\_tdist}$.
\\

\subsubsection{Water demand forecasting models}
The probabilistic water demand forecasts are constructed from a single univariate model, which is recursively applied in a Monte Carlo simulation. The obtained ensemble of trajectories approximates the process-specific multivariate distribution of the water demand process. 

\paragraph{LEAR-based model.}
The conditional distribution of water demand is specified as $W^{\mathrm{WD}}_t \sim \mathcal{N}(\mu_t, \sigma_t^2)$, where both $\mu_t$ and $\sigma_t^2$ depend on covariates. The corresponding predictors are estimated sequentially following \textcite{KleyHolsteg2020}: first, the conditional mean is fitted, and subsequently, the conditional variance is obtained by regressing on the squared residuals of the mean model. The location parameter is specified as
\begin{equation}
\begin{aligned}
\hat{\mu}_{t} =\;& \beta_{\mu,0} 
+ \sum_{i=1}^{336} \beta_{\mu,i}\, w^{\mathrm{WD}}_{t-i}
+ \sum_{i=3}^{8} \beta_{\mu,336+i-2}\, w^{\mathrm{WD}}_{t-168i} \\
&+ \sum_{i=1}^{24} \beta_{\mu,342+i}\, HoD_{i}
+ \sum_{i=1}^{168} \beta_{\mu,366+i}\, HoW_{i} \\
&+ HD(t) + HDI(D_{t-\mathcal{L}}, t) \\
&+ WF^{\text{as}}(t) + WF^{\text{sd}}(t) + WFI(t),
\end{aligned}
\label{eq:wdf_location}
\end{equation}
where $w^{\mathrm{WD}}_{t-i}$ denotes lagged realizations of the water demand of the last weeks. The variables $HoD_{i}$ and $HoW_{i}$ denote hour-of-day and hour-of-week dummy variables, respectively. Holiday effects are captured by $HD(t)$, where fixed-date and fixed-weekday holidays are distinguished. These effects are modeled by adjusting for weekday-specific patterns and by using cubic B-splines, following \textcite{Ziel2018}. The term $HDI(w^{\mathrm{WD}}_{t-\mathcal{L}}, t)$ represents interaction effects between $HD(t)$ and lagged demand, with $\mathcal{L} = \{1,24,168\}$. Weather effects are incorporated via $WF^{\text{as}}(t)$, $WF^{\text{sd}}(t)$, and $WFI(t)$. The component $WF^{\text{as}}(t)$ comprises aggregated statistics of weather forecast variables over predefined windows of 24 and 48 hours, including, for example, cumulative precipitation, mean temperature, and dry-period duration. The term $WF^{\text{sd}}(t)$ captures seasonal differences of the weather variables at a lag of 24 hours. Finally, $WFI(t)$ denotes a set of interaction features, including interactions within the aggregated feature space, within the weather variables, and between aggregated weather features and selected hour-of-day indicators. The variance model is specified as
\begin{equation}
\begin{aligned}
\hat{\sigma}^{2}_{t} =\;& \beta_{\sigma^2,0} 
+ \sum_{i=1}^{168} \beta_{\sigma^2,i}\, \hat{\varepsilon}_{t-i}^{2} 
+ \sum_{i=1}^{24} \beta_{\sigma^2,168 + i}\, HoD_{i} \\
&+ HD(t) 
+ WFI^{\text{HoD}}(t),
\end{aligned}
\label{eq:wdf_sigma2}
\end{equation}
where $\hat{\varepsilon}_{t-i}^{2}$ denotes the squared residuals of the mean model in Equation \eqref{eq:wdf_location}. The term $WFI^{\text{HoD}}(t)$ denotes a subset of $WFI(t)$ capturing interactions between aggregated weather features and selected hour-of-day indicators. To ensure non-negativity of the conditional variance, all coefficients are constrained to be non-negative, i.e., $\beta_{\sigma^2,\cdot} \geq 0$. We refer to this model as $\mathrm{WDF\_LEAR}$.

\paragraph{GAMLSS-based models.}
We consider specifications based on the normal distribution. The location predictor $\eta_{\mu,t}$ is modeled analogously to the conditional mean model in Equation \eqref{eq:wdf_location}. The predictor for $\eta_{\sigma,t}$ is specified by
\begin{equation}
\begin{aligned}
\eta_{\sigma,t} =\;& \beta_{\eta_{\sigma},0} 
+ \sum_{i = 1}^{168} \beta_{\eta_{\sigma},i}|w^{\mathrm{WD}}_{t-i}| 
+ \sum_{i=1}^{24} \beta_{\eta_{\sigma},168+i}\, HoD_{i}
+ WFI^{\text{HoD}}(t),
\end{aligned}
\label{eq:wdf_higher_moments}
\end{equation}
where we use absolute lagged observations $|w^{\mathrm{WD}}_{t-i}|$ as autoregressive information. The remaining inputs coincide with Equation \eqref{eq:wdf_sigma2} of the variance model of $\mathrm{WDF\_LEAR}$. We refer to the model as $\mathrm{WDF\_GAMLSS\_ndist}$.
\\

\subsubsection{Imbalance price forecasting models}
To model imbalance prices, we follow the water demand models, such that a single univariate model is estimated and then recursively applied in a Monte Carlo simulation. However, due to the pronounced volatility and occurrence of price spikes in imbalance prices, we apply additionally a median-normalized area hyperbolic sine (asinh) transformation with residual-based bias-corrected inversion to the input data, following \textcite{Narajewski2020}. Moreover, instead of modeling the final imbalance prices directly, the proposed framework considers their preliminary estimates, which are available immediately after delivery, whereas the final imbalance prices are published only after a substantial delay of several months. For notational simplicity, these preliminary estimates are likewise denoted as IBP.

\paragraph{LEAR-based model.}
The conditional mean of the transformed IBP process is specified as
\begin{equation}
\begin{aligned}
\hat{\mu}_{t} =\;& \beta_{\mu,0} 
+ \sum_{i=1}^{24} \beta_{\mu,i}\, \tilde{w}^{\mathrm{IBP}}_{t-i}
+ \beta_{\mu,25}\, \tilde{w}^{\mathrm{IBP}}_{t-168} \\
&+ \sum_{i=1}^{24} \beta_{\mu,49+i}\, HoD_{i}
+ \beta_{\mu,73}\, \tilde{w}^{\mathrm{DAP}}_{t} \\
&+ \text{Med}^{[t-168,t-1]}(\tilde{w}^{\mathrm{IBP}}),
\end{aligned}
\label{eq:ibpf_location}
\end{equation}
where $\tilde{w}^{\mathrm{IBP}}_{t}$ and $\tilde{w}^{\mathrm{DAP}}_{t}$ denote median-normalized asinh-transformed imbalance prices and day-ahead prices, respectively. The conditional variance is modeled as
\begin{equation}
\begin{aligned}
\hat{\sigma}^{2}_{t} =\;& \beta_{\sigma^2,0} 
+ \beta_{\sigma^2,1}\, \hat{\varepsilon}^{2}_{t-1}
+ \beta_{\sigma^2,2}\, \hat{\varepsilon}^{2}_{t-24}
+ \beta_{\sigma^2,3}\, \hat{\varepsilon}^{2}_{t-168} \\
&+ \sum_{i=1}^{24} \beta_{\sigma^2,3+i}\, HoD_{i}
+ \beta_{\sigma^2,29}\, \left|\tilde{w}^{\mathrm{DAP}}_{t}\right| \\
&+ \text{MAD}^{[t-168,t-1]}(\tilde{w}^{\mathrm{IBP}}),
\end{aligned}
\label{eq:ibpf_sigma2}
\end{equation}
where $\hat{\varepsilon}^{2}_{t}$ denotes the squared residuals of the transformed mean model in Equation \eqref{eq:ibpf_location}, and $|\tilde{w}^{\mathrm{DAP}}_{t}|$ denotes the absolute value of the transformed observed day-ahead price. For the mean and the variance model we used during model estimation, observed day-ahead prices. In the forecasting stage, simulated and subsequently transformed trajectories
\(\tilde{\hat{w}}^{DAP,(m)}_{h}\) are used.
This induces a dependence structure between the day-ahead and imbalance
price forecasts. The model is denoted as $\mathrm{IBPF\_LEAR}$.

\paragraph{GAMLSS-based models.}
As underlying distributions we consider also the Student’s t distribution to account for heavy tailed price data. The location predictor $\eta_{\mu,t}$ is specified analogously to the conditional mean model in equation \eqref{eq:ibpf_location}. The predictors for the remaining distributional parameters $\eta_{k,t}$, $k \neq \mu$, are given by
\begin{equation}
\begin{aligned}
\eta_{k,t} =\;& \beta_{k,0} 
+ \sum_{i=1}^{24} \beta_{k,i}\, HoD_{i}
+ \beta_{k,25}\, \left|\tilde{w}^{\mathrm{DAP}}_{t}\right| \\
&+ \text{MAD}^{[t-168,t-1]}(\tilde{w}^{\mathrm{IBP}}),
\end{aligned}
\label{eq:ibpf_higher_moments}
\end{equation}
where the covariates are specified in accordance with Equation \eqref{eq:ibpf_sigma2}, excluding the autoregressive terms. The corresponding model is denoted as $\mathrm{IBPF\_GAMLSS\_tdist}$.

\subsubsection{Joint forecast generation}
\begin{figure}[ht]
    \centering
    \includegraphics[width=\textwidth]{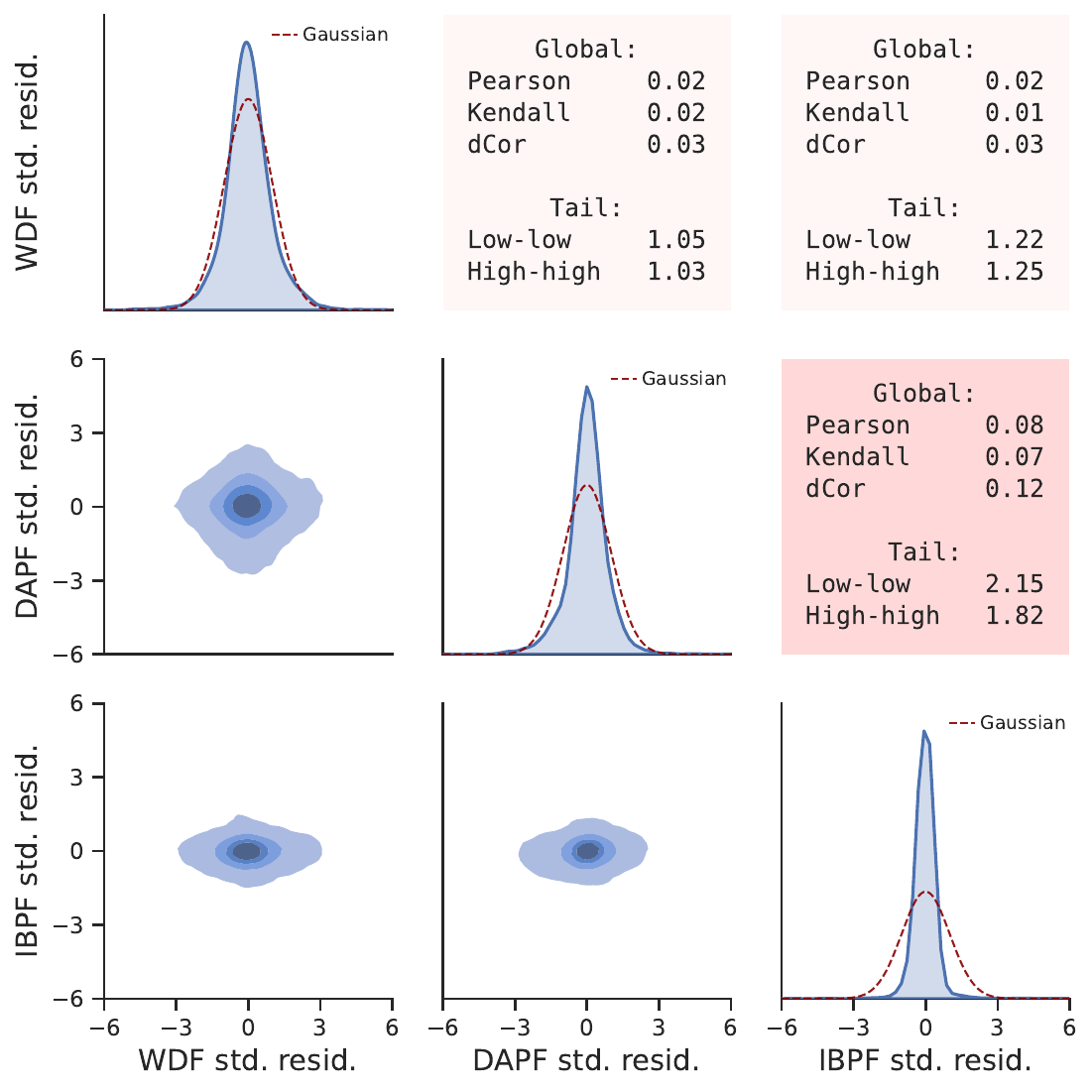}
    \caption{Illustration of the central and tail dependence structure between the standardized in-sample residuals of $\mathrm{WDF\_GAMLSS\_ndist}$, $\mathrm{DAPF\_GAMLSS\_tdist}$, and $\mathrm{IBPF\_GAMLSS\_tdist}$. The lower triangular panels show pairwise density plots of the corresponding processes, while the diagonal panels compare the marginal densities of the individual processes with a Gaussian reference density. The upper triangular panels report Pearson, Kendall, and distance correlation coefficients, representing linear, rank-based, and non-linear dependence structures, respectively. Tail dependence is assessed via parallel tail movement measures. Values close to \(1\) indicate independence, while larger values quantify the relative strength of joint tail movements compared to the independence case. The correlation structure is in general considered to be low, especially with respect to the pairing WDF-DAPF and WDF-IBPF.}
    \label{fig:residcorrelation}
\end{figure}

Given the three process-specific approximated multivariate forecast ensembles, a joint multivariate forecast distribution must be constructed that preserves the characteristics of each hourly marginal and the induced temporal dependency within each process. Methods such as the Schaake Shuffle and its extensions are well suited for this purpose \parencite{Clark2004}. However, an analysis of the pairwise contemporaneous correlations of the in-sample residuals across all three processes—both at the center and in the tails of the distributions—did not indicate sufficiently strong dependence as shown in Figure \ref{fig:residcorrelation}. We therefore adopt a simplifying assumption of independence between the water demand process and the electricity price processes. Joint trajectories are constructed via index-wise pairing, i.e., the $m$-th member of each ensemble is combined to form the $m$-th multivariate trajectory. This preserves the dependence structure between the day-ahead and imbalance price forecasts—induced through the modeling approach—while maintaining independence with respect to the water demand process. We refer to the obtained joint multivariate forecast ensembles as $\mathrm{Naive}$, $\mathrm{LEAR}$ and $\mathrm{GAMLSS}$.

\subsection{Policy models}
Since solving the base model directly is intractable, policy models are used to approximate the decision problem. As noted by \textcite{Powell2019}, the search over policy classes is recommended. However, for the sake of simplicity we restrict our attention to just one class of policies, namely Direct Lookahead Approximations. This class models explicitly how the system evolves along the optimization horizon given the exogenous information processes. We allow for the following approximations:

\paragraph{Reduction of decision variable space.}
To reduce the dimensionality of the decision space, a pump combination preselection heuristic is applied such that
\begin{equation}
x_t^{\mathrm{PC}} \in \mathcal{X}^{\mathrm{sel}} \subseteq \mathcal{X},
\qquad
t = 0, \dots, T,
\end{equation}
where \(\mathcal{X}^{\mathrm{sel}}\) denotes the subset of pump combinations considered within the optimization horizon. The preselection is initialized using the pump combination observed immediately before the optimization is started and follows a cluster-based heuristic with a predefined target size of required pump combinations. The heuristic balances the coverage of the feasible operating space with operational similarity to already selected pump combinations, where similarity reflects both the switching effort and the historically observed transition behavior. The candidate set is reduced from several hundred feasible pump combinations to nine. The target cardinality of nine corresponds to the 80th percentile of the number of distinct pump combinations observed over historical 36-hour time windows, matching the length of the optimization horizon.

\paragraph{MILP reformulation.}
To obtain a tractable MILP formulation, several logical and non-smooth components of the base model are reformulated using standard mixed-integer linear modeling techniques. First, big-M formulations as also applied in \textcite{Menke2016} are used to model conditional pump activation. In particular, flow and power variables are linked to binary pump operation variables such that they attain their predicted values when the corresponding pump combination is active and are forced to zero otherwise. Second, the CVaR and ER terms in Equations~\ref{eq:objective}
and~\ref{eq:reliabilitycontribution} are reformulated following
\textcite{Rockafellar2000}. As discussed by \textcite{Krokhmal2002},
scenario-based CVaR formulations can be transformed into linear programs
through the introduction of auxiliary variables. Any additional $(\cdot)^+$ operators arising from the proposed extensions of the ER formulations are linearized analogously.

\paragraph{Forecast generation and scenario reduction.}
Probabilistic forecasts are generated externally prior to the optimization and supplied as scenario inputs to the decision problem. As a result, the optimization model operates on forecast trajectories rather than on the underlying forecasting models, allowing the latter to be specified independently of the optimization framework.
The resulting joint multivariate forecast ensemble can be substantially larger than what is computationally tractable within the stochastic optimization framework. To reduce the computational burden, a scenario reduction procedure is therefore applied to the joint multivariate ensemble. Following \textcite{Ziel2021}, we employ a greedy forward selection algorithm with reweighting based on the Energy Distance (ED), as also adopted by \textcite{Heijden2025}. For two weighted ensembles \(X\) and \(X^*\), representing the full and reduced scenario sets with associated probability weights \(\omega\) and \(\omega^*\), respectively, the weighted empirical ED is given by
\begin{equation}
\begin{aligned}
ED_p(X, X^*)
=&\;
2 \sum_{i=1}^{M} \sum_{j=1}^{N}
\omega_i \omega^{*}_j
\, \|x^{(i)} - x^{*(j)}\|_2^{p}
\\
&-
\sum_{i=1}^{M} \sum_{j=1}^{M}
\omega_i \omega_j
\, \|x^{(i)} - x^{(j)}\|_2^{p}
\\
&-
\sum_{i=1}^{N} \sum_{j=1}^{N}
\omega^{*}_i \omega^{*}_j
\, \|x^{*(i)} - x^{*(j)}\|_2^{p},
\end{aligned}
\label{energydistance}
\end{equation}
where \(\|\cdot\|_2\) denotes the Euclidean norm and \(p \in (0,2)\) is the distance exponent, which determines the sensitivity of the distance measure to large pairwise deviations. The first term measures the discrepancy between the full and reduced ensembles, while the second and third terms quantify their respective within-ensemble dispersions. Since the dispersion term of the full ensemble \(X\) remains constant during the selection procedure, it does not affect the optimization and can therefore be omitted. Notably, the resulting objective is closely related to the Energy Score (ES), a proper scoring rule commonly used for the evaluation of multivariate probabilistic forecasts.

\begin{figure}[ht]
    \centering
    \includegraphics[
    width=\textwidth,
    height=0.8\textheight,
    keepaspectratio
]{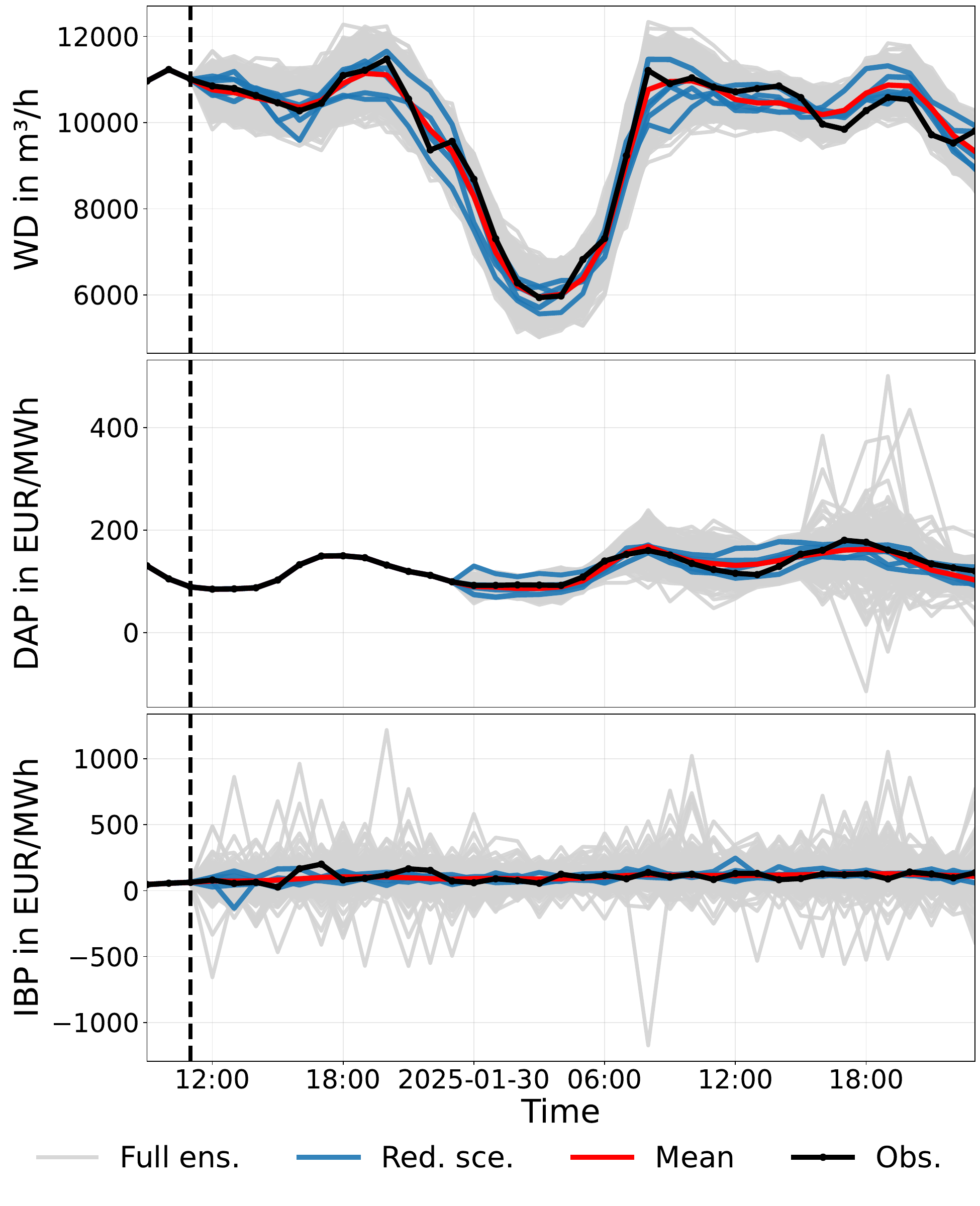}
    \caption{Illustration of full ensemble with mean as well as reduced scenario set forecasts of the $\mathrm{WDF\_GAMLSS\_ndist}$, $\mathrm{DAPF\_GAMLSS\_tdist}$, and $\mathrm{IBPF\_GAMLSS\_tdist}$ models compared to their corresponding observations.}
    \label{fig:fullredscenarios}
\end{figure}

The forward selection procedure iteratively constructs the reduced ensemble by successively adding ensemble members that yield the lowest score in the reduced ED. In addition to selecting a representative subset \(X^* \subset X\), the associated probability weights \(\omega^*\) are optimized to best approximate the full distribution \(X\). Following \textcite{Ziel2021}, the weights are obtained by solving the quadratic program
\begin{equation}
\omega^{*}
=
\arg\min_{\omega^{*}} \; ED_p(X, X^*),
\quad
\text{s.t.} \quad \sum_{j=1}^{N} \omega^{*}_j = 1, \quad \omega^{*}_j \geq 0.
\end{equation}
We determined the number of retained scenarios as \(M^{\mathrm{red}}=5\) and set the distance exponent to \(p=1\). The resulting reduced scenario set serves as input to the stochastic optimization framework, as illustrated in Figure~\ref{fig:fullredscenarios}. Scenario reduction is particularly challenging in the present setting for three reasons. First, the joint multivariate ensemble combines water demand, day-ahead prices, and imbalance prices, which exhibit markedly different temporal dynamics and distributional characteristics, even after standardization. Second, reducing the full ensemble to only five scenarios requires a substantial compression of the underlying uncertainty representation. Third, the optimization problem is risk-driven, implying that an accurate representation of tail events may be more important than an accurate approximation of the overall probability distribution. As discussed by \textcite{Arpon2018}, scenarios associated with extreme outcomes can have a disproportionate impact on risk-averse decision making.

\begin{figure}[ht]
    \centering
    \includegraphics[width=\textwidth]{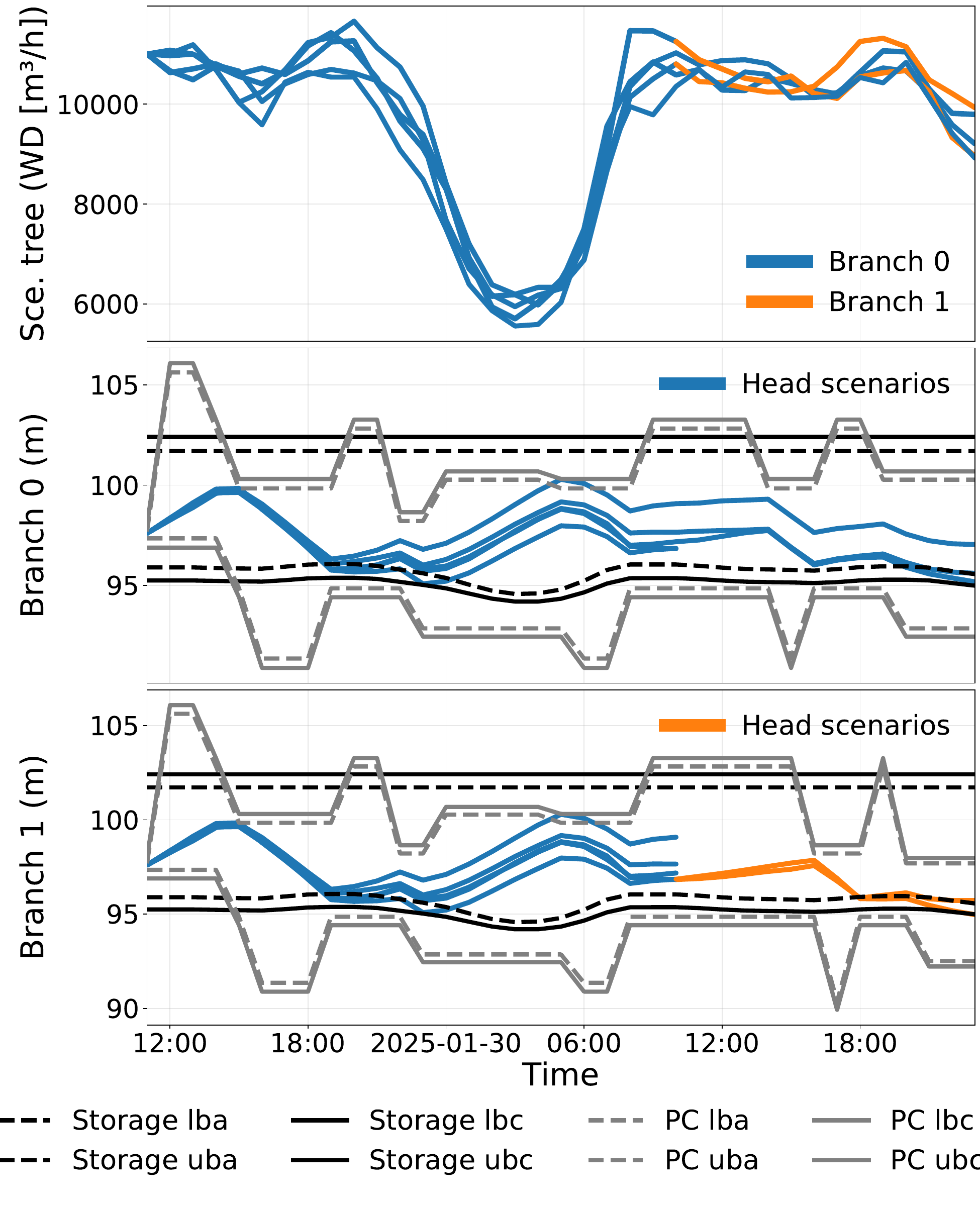}
    \caption{Scenario tree representation based on WD process and the corresponding branches within the stochastic optimization policy for $\mathrm{GAMLSS\_SH}$.}
    \label{fig:scetree}
\end{figure}

\paragraph{Scenario tree generation.}
The reduced scenario set is represented as a multistage scenario tree \parencite{Heitsch2009}. To limit the complexity of the scenario tree, only the water demand process is considered for branching, as it primarily determines the reliable operation of the physical system. Specifically, branching is triggered whenever the spread of cumulative water-demand trajectories within a branch exceeds a threshold derived from the feasible storage-volume and pump-head ranges associated with the preselected pump combinations, as illustrated in Figure~\ref{fig:scetree}. Although the scenario tree defines the branching structure for the sequential decision process, risk measures are evaluated on the original reduced scenario set. Consequently, quantities such as CVaR are computed across all reduced scenarios rather than within individual branches.

\paragraph{Sequential overlapping horizon optimization.}
To ensure computational tractability, the full optimization horizon is decomposed into a sequence of overlapping sub-horizons. Specifically, each sub-horizon spans 12 hours, with an overlap of 6 hours between consecutive problems. The sub-horizons are solved sequentially under a fixed information set. The solution of each sub-horizon provides the initial conditions for the subsequent problem, ensuring consistency across the full horizon.
\\

\subsubsection{Forecast-based policy models}
Based on the joint multivariate forecast ensembles introduced as $\mathrm{Naive}$, $\mathrm{LEAR}$, and $\mathrm{GAMLSS}$, we construct different policy models denoted as $\mathrm{Naive\_SH}$, $\mathrm{LEAR\_SH}$ and $\mathrm{GAMLSS\_SH}$, where $\mathrm{SH}$ denotes the sequential overlapping horizon optimization without information update. Moreover, we also introduce deterministic ($\mathrm{DET}$) versions denoted as $\mathrm{Naive\_DET\_FH}$, a $\mathrm{LEAR\_DET\_FH}$, and a $\mathrm{GAMLSS\_DET\_FH}$. In contrast to their stochastic counterparts, they are solved in a one shot procedure along the full optimization horizon denoted as $\mathrm{FH}$. The stochastic MILP policy framework reduces to a deterministic MILP when the exogenous processes are represented by deterministic inputs.

\subsubsection{Perfect foresight policy models}
To provide a lower bound for the underlying base model approximation, we provide the three exogenous processes as oracle forecasts. Here, we introduce two versions, to highlight the price for complexity reduction due to decomposition of the problem in sub horizons. We refer to the policy models as $\mathrm{PF\_DET\_SH}$ and $\mathrm{PF\_DET\_FH}$, respectively.

\subsubsection{Fall-back policy model}
The fall-back policy model is designed to approximate conventional operator behavior in the absence of dynamic price incentives. Consequently, the policy focuses primarily on secure and efficient system operation. The fall-back policy serves two purposes. First, it provides an operational benchmark. Second, it acts as a safety mechanism during policy evaluation to better reflect practical operating procedures, where human operators would intervene once critical operating conditions are reached. The fall-back policy is constructed according to the following operational principles. Since storage facilities have traditionally been replenished during night-time hours, the admissible lower storage bound is set to $v_t^{\mathrm{lba}} = 30,000$ for the early morning hours $t \in \{14,15,16\}$, corresponding to approximately $68\%$ of the physical storage capacity. This encourages replenishment of the storage tank during the night and establishes a reserve for subsequent daytime operation. To eliminate dynamic price incentives, the DAP process is replaced by the average DAP observed during the training period. Furthermore, the non-anticipativity condition for the water demand process is relaxed by assuming perfect foresight. This reflects the operator's ability to anticipate recurring demand patterns from operational experience. The resulting benchmark policy is denoted by $\mathrm{FB\_DET\_FH}$.

\subsection{Forecast and policy evaluation framework}
To evaluate both the policy models and the forecasting models we conduct a numerical simulation study, where randomly $N^{\mathrm{runs}}=100$ runs are drawn from the test data set with each $N^{\mathrm{reps}}=3$ repetitions. 

\subsubsection{Forecast evaluation}
Since the joint forecast ensemble is obtained by coupling separately generated forecast distributions, forecast quality is evaluated from both a process-specific and a joint perspective. To this end, we employ proper scoring rules \parencite{Gneiting2007} to assess point forecasts, marginal forecasts and process-specific multivariate forecast distributions, and the joint multivariate forecast distribution spanning all processes. Let $y_t$ denote the observed realization at time step $t=1,\dots,T$, $\hat{y}_t$ the corresponding point forecast, and $x_t^{(m)}$ the increment of the $m$-th ensemble member at time step $t$, where $m=1,\dots,M$. Here, $T$ denotes the forecast horizon length and $M$ the ensemble size. For both multivariate evaluations, we define the forecast and observation trajectories as
\[
x^{(m)} = (x^{(m)}_1,\dots,x^{(m)}_T)
\quad\text{and}\quad
y = (y_1,\dots,y_T).
\]
Depending on the evaluation setting, these vectors either represent the trajectory of a single process (process-specific evaluation) or the standardized trajectories of all processes (joint evaluation).
Point forecast accuracy is evaluated using the mean absolute error (MAE),

\begin{equation}
\mathrm{MAE}
=
\frac{1}{T}
\sum_{t=1}^{T}
\left|
y_t - \hat{y}_t
\right|,
\end{equation}
and the root mean squared error (RMSE),
\begin{equation}
\mathrm{RMSE}
=
\sqrt{
\frac{1}{T}
\sum_{t=1}^{T}
\left(
y_t - \hat{y}_t
\right)^2
}.
\end{equation}
Marginal probabilistic performance is evaluated using the continuous ranked probability score (CRPS) to evaluate the full marginal distribution,
\begin{equation}
\mathrm{CRPS}
=
\frac{1}{T}
\sum_{t=1}^{T}
\left[
\frac{1}{M}
\sum_{m=1}^{M}
\left|x^{(m)}_{t}-y_t\right|
-
\frac{1}{2M^2}
\sum_{m=1}^{M}
\sum_{j=1}^{M}
\left|x^{(m)}_{t}-x^{(j)}_{t}\right|
\right],
\end{equation}
To assess the accuracy of specific quantile levels, we employ the pinball score (PB). As the optimization framework is driven by CVaR- and ER-based risk measures, which primarily depend on extreme realizations, particular emphasis is placed on the tail regions of the forecast distributions. The PB is defined as

\begin{equation}
\mathrm{PB}
=
\frac{1}{|\mathcal{T}|}
\sum_{\tau \in \mathcal{T}}
\left[
\frac{1}{T}
\sum_{t=1}^{T}
\left(
\mathbf{1}_{\{y_t < q_{\tau,t}\}}
-
\tau
\right)
\left(
q_{\tau,t}-y_t
\right)
\right],
\end{equation}

where \(\mathcal{T}=\{0.01,0.02,\ldots,0.99\}\) denotes the set of evaluated quantile levels and \(q_{\tau,t}\) the predicted \(\tau\)-quantile at time step \(t\). The lower and upper tail regions are defined as

\begin{equation}
\mathcal{T}^{\mathrm{low}}
=
\{\tau \in \mathcal{T} : \tau \leq 0.20\},
\qquad
\mathcal{T}^{\mathrm{up}}
=
\{\tau \in \mathcal{T} : \tau \geq 0.80\}.
\end{equation}

For \(\hat{w}^{\mathrm{WD}}\) and \(\hat{w}^{\mathrm{IBP}}\), both lower and upper tails are especially decision relevant. For \(\hat{w}^{\mathrm{DAP}}\), only the upper tail is considered. In the following, \(\mathcal{T}^{\mathrm{tail}}\) denotes the corresponding tail quantile set under consideration.
To assess the multivariate forecast distribution the ES is applied,
\begin{equation}
\mathrm{ES}
=
\frac{1}{M}
\sum_{m=1}^{M}
\left\|
x^{(m)}-y
\right\|^p
-
\frac{1}{2M^2}
\sum_{m=1}^{M}
\sum_{j=1}^{M}
\left\|
x^{(m)}-x^{(j)}
\right\|^p,
\end{equation}
where the exponent is set to \(p=1\). As the accurate representation of temporal dependence is crucial in storage optimization due to the intertemporal coupling of decisions and system states, we additionally employ the Variogram Score (VS) to explicitly assess the preservation of the temporal dependence structure for the process-specific and joint multivariate distribution.
\begin{equation}
\mathrm{VS}
=
\frac{2}{T(T-1)}
\sum_{i=1}^{T-1}
\sum_{j=i+1}^{T}
\left(
\frac{1}{M}
\sum_{m=1}^{M}
\left|
x_i^{(m)} - x_j^{(m)}
\right|^{p}
-
\left|
y_i-y_j
\right|^{p}
\right)^2,
\label{eq:variogram_score}
\end{equation}
where $p$ denotes the variogram order, which is set to $p=1$. For comparability across forecast horizons and processes, the variogram score was additionally normalized by the squared trajectory dimension and square-root transformed during implementation. 
For evaluation of the weighted reduced scenario sets, the probabilistic scoring rules are extended to their weighted versions. Statistical differences in predictive accuracy are assessed using the Diebold--Mariano (DM) test \parencite{Diebold1995,Diebold2002}. The test compares the mean loss differentials between two competing forecasting models based on the computed scalar losses. Rejection of the null hypothesis indicates statistically significant differences in predictive performance between the compared models. The scalar losses are obtained by averaging the corresponding repetitions within the numerical study per time point. 

\subsubsection{Policy evaluation}
The policies are evaluated using realized trajectories of the exogenous processes, including the realized and not estimated imbalance prices for imbalance settlement. The proposed fall-back policy is applied as an emergency strategy. Following the first violation of secure system operation beyond lower and upper critical operating bounds, the emergency strategy remains active for the remainder of the optimization horizon. The resulting intervention costs are quantified through imbalance costs arising from deviations in pumping power between the procurement of the corresponding policy and the subsequently applied fall-back policy. The policies are evaluated using the following performance indicators:

\begin{enumerate}
\renewcommand{\labelenumi}{\roman{enumi})}

\item \textbf{Economical costs:} They measure the economic costs associated with a policy. \emph{Norm. costs (EUR)} denotes the normalized total operational costs, consisting of day-ahead procurement and imbalance costs. \emph{Imb. costs (EUR)} reports the imbalance costs separately to quantify the impact of balancing actions. Furthermore, \emph{Share abs. imb. energy} measures the relative share of absolute imbalance energy.

\item \textbf{System operation reliability:} They measure the sustainability and realiability associated with a specific policy. \emph{Norm. energy (MWh)} denotes the normalized total pumping energy. Furthermore, \emph{First violation (h)} reports the average first-hour violation of system reliability constraints and \emph{Avg. contin. pump runtime (h)} the average number of consecutive pump operating hours.

\item \textbf{Computational Performance:} These metrics assess the computational effort and solution quality achieved during optimization. \emph{Relative gap} reports the average MILP optimality gap at termination, while \emph{Time (s)} denotes the average optimization runtime including the preprocessing and forecast generation time in seconds.

\end{enumerate}

The quantities \emph{Norm. costs (EUR)}, \emph{Norm. energy (MWh)}, \emph{Share abs. imb. energy}, and \emph{Imb. costs (EUR)} are evaluated only over the restricted period from 00{:}00 to 23{:}00 per optimization run. The beginning hours are excluded to account for the retroactive day-ahead participation assumption (Assumption~\ref{ass:studydesign}). The normalized quantities are normalized by the pumped water volume obtained under the $\mathrm{PF\_DET\_FH}$ policy to improve comparability. All reported values correspond to averages over the corresponding runs and repetitions.\\

The resulting MILP problems are solved using the \texttt{Gurobi} optimizer \parencite{Gurobi2025}. The solver settings include a relative optimality gap of $1\%$, an absolute optimality gap of 1 EUR, and a feasibility tolerance of ($10^{-6}$). A time limit of 20 minutes is imposed on each solver run. The numerical study were performed on an Intel Core i7-11370H workstation with 32 GB RAM using eight parallel solver threads.

\section{Results and Discussion}
In this section, we present and discuss the results of the numerical study. We first assess the quality of the probabilistic forecasts, then discuss the scenario reduction approach, and finally present the performance of the resulting policy models.

\subsection{Forecast performance}
\begin{figure}[ht]
    \centering
    \includegraphics[width=\textwidth]{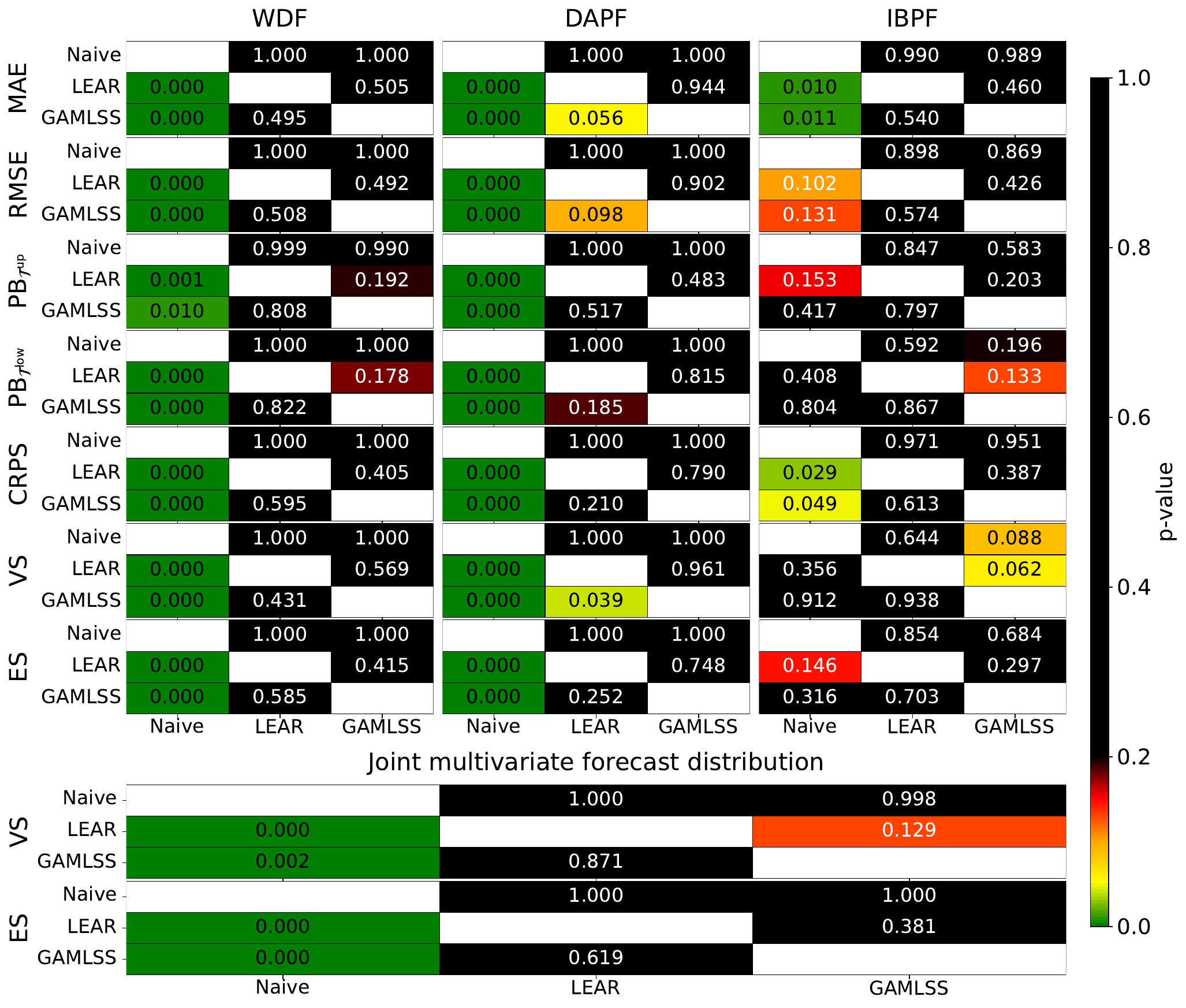}
    \caption{DM test results for reduced scenario sets derived from the forecasting models underlying $\mathrm{Naive\_SH}$, $\mathrm{LEAR\_SH}$, and $\mathrm{GAMLSS\_SH}$. The labels $\mathrm{Naive}$, $\mathrm{LEAR}$, and $\mathrm{GAMLSS}$ refer to the corresponding joint multivariate forecast distributions.}
    \label{fig:dmtest}
\end{figure}

Table \ref{tab:results_forecast} summarizes the forecast performance of the reduced and full scenario sets, while Figure \ref{fig:dmtest} reports the corresponding significance tests for the reduced sets. Overall, water demand and day-ahead price forecasts substantially outperform imbalance price forecasts, both in absolute terms and relative to the naive benchmark. For water demand, LEAR exhibits a slight but generally not statistically significant advantage over GAMLSS, whereas GAMLSS performs marginally better for day-ahead prices. In contrast, all approaches show comparatively poor performance for imbalance prices. Notably, the reduced scenario set of $\mathrm{GAMLSS\_tdist}$ is outperformed by the naive benchmark for some metrics, while the corresponding full scenario set is not. At the joint level, LEAR and GAMLSS achieve nearly identical performance and clearly outperform the naive benchmark, suggesting only modest differences between the advanced forecasting approaches.

\subsection{Scenario reduction performance}
\begin{figure}[ht]
    \centering
    \includegraphics[
      width=\textwidth,
      height=0.7\textheight,
      keepaspectratio
    ]{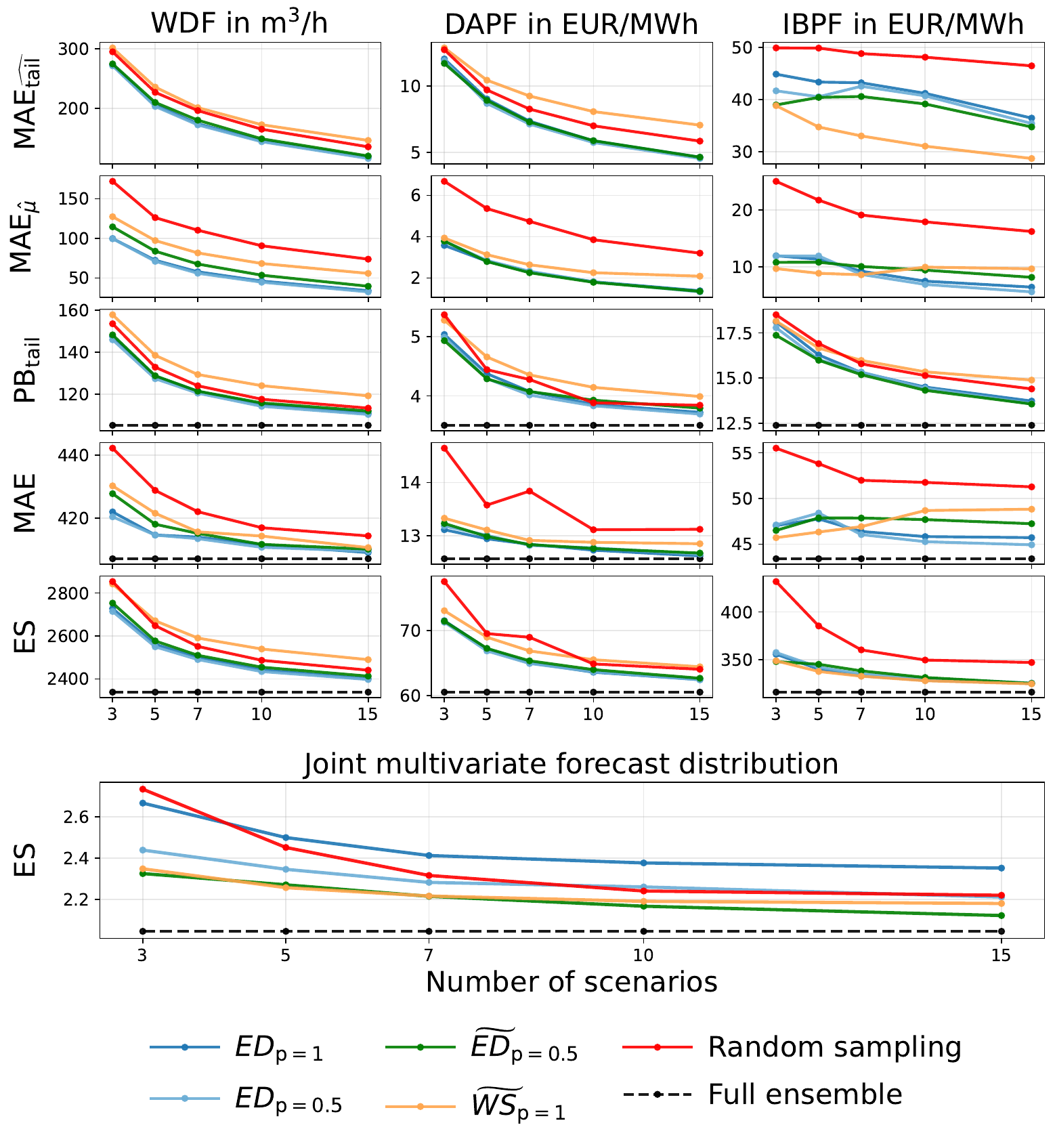}
    \caption{Comparison of scenario reduction methods for the $\mathrm{GAMLSS}$ based models for the Energy Distance (ED), Wasserstein Distance (WS), and Random Sampling using a forward selection algorithm for varying set sizes of scenario members. The symbol $\tilde{\cdot}$ denotes the corresponding methods applied after a median-normalized asinh transformation. The two upper panel rows evaluate the ability of the reduced scenario sets to reproduce the statistical properties of the full ensemble in terms of tail quantiles and mean values. The following three panal rows assess the out-of-sample predictive performance of both the reduced and full scenario sets against the observed realizations. The plot in the last rows corresponds to the joint multivariate performance.}
    \label{fig:scered}
\end{figure}

To assess scenario reduction performance, we compare the proposed ED with the Wasserstein distance (WS) \parencite{Dupacova2003} and random sampling using the \(\mathrm{GAMLSS}\)-based forecasts (Figure~\ref{fig:scered}). Since scenario reduction is performed jointly across all forecasts, we consider both standardization and a median-normalized asinh transformation to account for heterogeneous scales and tail behavior. In addition, an ED with \(p=0.5\) is evaluated to reduce the influence of extreme observations \parencite{Ziel2021}. The reduced scenario sets are assessed with respect to their ability to reproduce the full ensemble and their out-of-sample predictive performance at both the process-specific and joint levels.
Overall, WS achieves results comparable to ED only after applying the median-normalized asinh transformation, whereas ED exhibits comparatively stable performance across preprocessing variants. Nevertheless, its joint predictive performance benefits from the transformation and the reduced emphasis on extreme observations induced by \(p=0.5\).
A notable finding is the mismatch between the scenario reduction and the downstream optimization objective. To investigate this, we evaluate tail preservation using the proposed metrics $\mathrm{MAE}_{\widehat{\mathrm{tail}}}$ and $\mathrm{PB}_{\mathrm{tail}}$. While $\mathrm{MAE}_{\widehat{\mathrm{tail}}}$ quantifies how well the tail regions of the full scenario distribution are preserved after scenario reduction, $\mathrm{PB}_{\mathrm{tail}}$ evaluates the out-of-sample predictive accuracy of the reduced scenarios for observations in the tail regions. Under these measures, random sampling proves surprisingly competitive. While ED- and WS-based methods are designed to approximate the overall distribution, they do not explicitly prioritize tail events \parencite{Ziel2021}. Consequently, a better approximation of the full distribution does not necessarily imply better preservation of the tail characteristics relevant to the CVaR- and ER-based optimization problem.

\subsection{Policy performance}
After evaluating the forecasting and scenario reduction approaches, we turn to the resulting policy models summarized in Table~\ref{tab:results_opt}. Here, \(\mathrm{FB\_DET\_FH}\) and \(\mathrm{PF\_DET\_FH}\) provide upper and lower performance bounds for the sequential decision problem.
For the forecast-derived policies, the ranking largely mirrors the forecasting performance of the underlying models. Although deterministic and stochastic variants achieve similar normalized costs, deterministic policies incur substantially more imbalance energy and exhibit reduced operational robustness, reflected by earlier first-hour violations. This suggests that the primary benefit of stochastic optimization lies in improved risk management and system reliability rather than lower expected costs. Using \(\mathrm{GAMLSS\_SH}\) as a reference, we perform a sensitivity analysis.

With respect to economic risk aversion, purely expectation-driven (\(\lambda^{\mathrm{risk}}=1\)) and purely risk-driven (\(\lambda^{\mathrm{risk}}=0\)) formulations yield only minor differences. Several factors may contribute to this observation. The scenario reduction procedure is not explicitly designed to preserve tail events, reliability requirements substantially restrict the operational flexibility available for managing economic risk, imbalance prices exhibit limited predictability, and global decision constraints partially align expectation- and risk-driven policies.

Regarding the scenario reduction procedure, increasing the scenario set size to \(M^{\mathrm{red}}=10\) and employing the transformed Energy Distance \(\widetilde{\mathrm{ED}}_{0.5}\) each yield modest performance improvements, indicating limited but positive benefits from a richer representation of uncertainty.

Finally, to assess the value of imbalance price forecasts, we compare two cost-only penalty formulations: one using forecasted imbalance prices and one using a constant penalty corresponding to the \(0.99\)-quantile of historical imbalance prices. Both formulations further reduce the absolute imbalance share. However, the forecast-based formulation achieves superior cost performance, indicating that imbalance price forecasts still provide economically valuable information. Likewise, \(\mathrm{GAMLSS\_PF\_EPF\_SH}\), which employs perfect day-ahead and imbalance price forecasts, yields only moderate improvements over its forecast-based counterpart.
\\

Overall, the results are encouraging. Taking \(\mathrm{FB\_DET\_FH}\) as a benchmark and assuming an imbalance risk premium of \(2\,\mathrm{EUR/MWh}\), the best-performing policy, \(\mathrm{LEAR\_SH}\), yields extrapolated annual cost savings of \(236{,}355.75\,\mathrm{EUR}\), corresponding to approximately \(9\,\%\). This indicates that substantial economic benefits can be realized by accepting a carefully controlled increase in imbalance exposure and operational risk.

More generally, the results suggest that improvements in forecasting performance translate into improved policy performance. Stochastic policies consistently reduce imbalance volumes and improve operational robustness, while the sensitivity analysis indicates that reliability constraints and water demand uncertainty are the dominant drivers of the optimization problem. Although electricity price forecasts are economically relevant, further improvements translate only marginally into policy performance, suggesting that most exploitable price information is already captured.

Future research may consider renewable energy integration and multi-market participation. From a methodological perspective, promising directions include decision-aware scenario reduction, alternative policy classes, and receding-horizon MPC formulations with an explicit handling of forecast updates, since subsequent optimization runs may revise decisions based on improved forecasts \parencite{Ghadimi2024}.

\begin{table}[ht]
\centering
\tiny
\renewcommand{\arraystretch}{1.7}
\setlength{\tabcolsep}{1.5pt}
\caption{Performance of reduced forecast scenarios with full forecast ensemble results in brackets.}
\label{tab:results_forecast}
\begin{adjustbox}{max width=\textwidth}
\input{scores_forecast_evaluation}
\end{adjustbox}
\end{table}

\begin{table}[ht]
\centering
\tiny
\renewcommand{\arraystretch}{1.7}
\setlength{\tabcolsep}{1pt}
\begin{adjustbox}{max width=\textwidth}
\input{opt_results_table_heatmap}
\end{adjustbox}
\vspace{1mm}
\begin{minipage}{0.9\textwidth}
\footnotesize
$^{*}$ Absolute $\mathrm{IBPF}$ and imbalance volumes (cost-only penalty).\\
$^{**}$ $0.99$-quantile of absolute historical $\mathrm{IBP}$ and absolute imbalance volumes (cost-only penalty).
\end{minipage}
\caption{Performance of policy models.}
\label{tab:results_opt}
\end{table}

\section{Summary}
The study addressed a sequential pump scheduling problem under uncertainty by combining probabilistic forecasts of water demand and electricity prices with a risk-aware stochastic optimization framework. The generalized formulation enhances transferability across water suppliers and can be readily extended to incorporate site-specific hydraulic models and operational constraints. Particular emphasis was placed on the integration of forecasting and optimization, the explicit treatment of operational and economic risks, and the modeling of imbalance market settlement.

The results show that improvements in forecasting performance generally translate into improved policy performance. They further highlight the importance of scenario reduction and suggest that, when tail-sensitive risk measures such as CVaR and ER are employed, scenario reduction methods should place greater emphasis on preserving decision-relevant tail behavior to better align reduced scenario sets with the downstream optimization objectives.

Furthermore, the primary benefit of stochastic optimization lied in improved operational robustness rather than lower expected costs. Compared with deterministic approaches, stochastic policies reduced imbalance energy, delayed operational constraint violations, and improved overall system reliability.

\section{Funding}
The authors declare that no funds, grants, or other support were received during the preparation of this manuscript.

\section{Competing Interests}
The authors have no relevant financial or non-financial interests to disclose.

\section{Generative AI Statement}
During the preparation of this manuscript, generative artificial intelligence (AI) tools have been used, including ChatGPT and Cursor AI, to support software development, discuss methodological approaches, improve the structure and clarity of the manuscript, and assist with code refinement. All generated text and code suggestions were critically reviewed, verified, and revised by the authors. The authors are fully responsible for the scientific content, methodology, analyses, interpretations, and conclusions presented in this manuscript.

\clearpage
\printbibliography
\end{document}

%% file: scores_forecast_evaluation.tex
\begin{tabular}{cccccccc}
\toprule
 & MAE & RMSE & $PB_{\mathcal{T}^{\mathrm{up}}}$ & $PB_{\mathcal{T}^{\mathrm{low}}}$ & CRPS & VS & ES \\
\midrule
\textbf{WDF} & \bfseries {\cellcolor{white}}  & \bfseries {\cellcolor{white}}  & \bfseries {\cellcolor{white}}  & \bfseries {\cellcolor{white}}  & \bfseries {\cellcolor{white}}  & \bfseries {\cellcolor{white}}  & \bfseries {\cellcolor{white}}  \\
$\mathrm{LEAR}$ & {\cellcolor[HTML]{006837}} \color[HTML]{F1F1F1} 413.5 (402.8) & {\cellcolor[HTML]{006837}} \color[HTML]{F1F1F1} 525.9 (512.6) & {\cellcolor[HTML]{006837}} \color[HTML]{F1F1F1} 125.5 (97.4) & {\cellcolor[HTML]{006837}} \color[HTML]{F1F1F1} 109.4 (92.3) & {\cellcolor[HTML]{006837}} \color[HTML]{F1F1F1} 327.4 (293.4) & {\cellcolor[HTML]{07753E}} \color[HTML]{F1F1F1} 581.9 (559.3) & {\cellcolor[HTML]{006837}} \color[HTML]{F1F1F1} 2524.3 (2257.9) \\
$\mathrm{GAMLSS\_ndist}$ & {\cellcolor[HTML]{006837}} \color[HTML]{F1F1F1} 413.2 (406.9) & {\cellcolor[HTML]{006837}} \color[HTML]{F1F1F1} 526.5 (518.6) & {\cellcolor[HTML]{4EB15D}} \color[HTML]{F1F1F1} 139.7 (108.4) & {\cellcolor[HTML]{33A456}} \color[HTML]{F1F1F1} 120.3 (98.2) & {\cellcolor[HTML]{05713C}} \color[HTML]{F1F1F1} 332.5 (304.1) & {\cellcolor[HTML]{006837}} \color[HTML]{F1F1F1} 576.9 (563.7) & {\cellcolor[HTML]{06733D}} \color[HTML]{F1F1F1} 2560.7 (2338.5) \\
$\mathrm{Naive}$ & {\cellcolor[HTML]{A50026}} \color[HTML]{F1F1F1} 670.9 (658.6) & {\cellcolor[HTML]{A50026}} \color[HTML]{F1F1F1} 813.0 (797.5) & {\cellcolor[HTML]{A50026}} \color[HTML]{F1F1F1} 208.1 (168.4) & {\cellcolor[HTML]{A50026}} \color[HTML]{F1F1F1} 191.0 (154.0) & {\cellcolor[HTML]{A50026}} \color[HTML]{F1F1F1} 546.0 (487.7) & {\cellcolor[HTML]{A50026}} \color[HTML]{F1F1F1} 748.2 (725.2) & {\cellcolor[HTML]{A50026}} \color[HTML]{F1F1F1} 3991.9 (3573.8) \\
\addlinespace
\midrule
\addlinespace
\textbf{DAPF} & \bfseries {\cellcolor{white}}  & \bfseries {\cellcolor{white}}  & \bfseries {\cellcolor{white}}  & \bfseries {\cellcolor{white}}  & \bfseries {\cellcolor{white}}  & \bfseries {\cellcolor{white}}  & \bfseries {\cellcolor{white}}  \\
$\mathrm{LEAR\_QRA}$ & {\cellcolor[HTML]{1E9A51}} \color[HTML]{F1F1F1} 14.5 (14.1) & {\cellcolor[HTML]{1E9A51}} \color[HTML]{F1F1F1} 19.1 (18.5) & {\cellcolor[HTML]{006837}} \color[HTML]{F1F1F1} 4.3 (3.5) & {\cellcolor[HTML]{108647}} \color[HTML]{F1F1F1} 4.0 (3.2) & {\cellcolor[HTML]{0E8245}} \color[HTML]{F1F1F1} 11.2 (10.0) & {\cellcolor[HTML]{75C465}} \color[HTML]{000000} 22.5 (21.9) & {\cellcolor[HTML]{0E8245}} \color[HTML]{F1F1F1} 71.5 (63.7) \\
$\mathrm{GAMLSS\_tdist}$ & {\cellcolor[HTML]{006837}} \color[HTML]{F1F1F1} 12.9 (12.6) & {\cellcolor[HTML]{006837}} \color[HTML]{F1F1F1} 17.1 (16.7) & {\cellcolor[HTML]{016A38}} \color[HTML]{F1F1F1} 4.3 (3.4) & {\cellcolor[HTML]{006837}} \color[HTML]{F1F1F1} 3.6 (2.9) & {\cellcolor[HTML]{006837}} \color[HTML]{F1F1F1} 10.4 (9.4) & {\cellcolor[HTML]{006837}} \color[HTML]{F1F1F1} 19.6 (19.2) & {\cellcolor[HTML]{006837}} \color[HTML]{F1F1F1} 67.2 (60.5) \\
$\mathrm{Naive}$ & {\cellcolor[HTML]{A50026}} \color[HTML]{F1F1F1} 28.6 (27.5) & {\cellcolor[HTML]{A50026}} \color[HTML]{F1F1F1} 35.7 (34.8) & {\cellcolor[HTML]{A50026}} \color[HTML]{F1F1F1} 8.7 (7.5) & {\cellcolor[HTML]{A50026}} \color[HTML]{F1F1F1} 9.0 (7.4) & {\cellcolor[HTML]{A50026}} \color[HTML]{F1F1F1} 23.8 (21.3) & {\cellcolor[HTML]{A50026}} \color[HTML]{F1F1F1} 32.8 (32.4) & {\cellcolor[HTML]{A50026}} \color[HTML]{F1F1F1} 141.8 (127.3) \\
\addlinespace
\midrule
\addlinespace
\textbf{IBPF} & \bfseries {\cellcolor{white}}  & \bfseries {\cellcolor{white}}  & \bfseries {\cellcolor{white}}  & \bfseries {\cellcolor{white}}  & \bfseries {\cellcolor{white}}  & \bfseries {\cellcolor{white}}  & \bfseries {\cellcolor{white}}  \\
$\mathrm{LEAR}$ & {\cellcolor[HTML]{006837}} \color[HTML]{F1F1F1} 45.0 (43.2) & {\cellcolor[HTML]{006837}} \color[HTML]{F1F1F1} 64.5 (62.5) & {\cellcolor[HTML]{006837}} \color[HTML]{F1F1F1} 13.7 (11.3) & {\cellcolor[HTML]{006837}} \color[HTML]{F1F1F1} 15.6 (11.0) & {\cellcolor[HTML]{006837}} \color[HTML]{F1F1F1} 37.1 (33.0) & {\cellcolor[HTML]{006837}} \color[HTML]{F1F1F1} 73.6 (69.1) & {\cellcolor[HTML]{006837}} \color[HTML]{F1F1F1} 315.8 (288.8) \\
$\mathrm{GAMLSS\_tdist}$ & {\cellcolor[HTML]{0A7B41}} \color[HTML]{F1F1F1} 45.3 (43.4) & {\cellcolor[HTML]{3FAA59}} \color[HTML]{F1F1F1} 66.0 (62.6) & {\cellcolor[HTML]{ED5F3C}} \color[HTML]{F1F1F1} 14.8 (11.8) & {\cellcolor[HTML]{A50026}} \color[HTML]{F1F1F1} 18.0 (12.7) & {\cellcolor[HTML]{3FAA59}} \color[HTML]{F1F1F1} 38.0 (34.4) & {\cellcolor[HTML]{A50026}} \color[HTML]{F1F1F1} 96.0 (83.1) & {\cellcolor[HTML]{FFF2AA}} \color[HTML]{000000} 340.3 (316.2) \\
$\mathrm{Naive}$ & {\cellcolor[HTML]{A50026}} \color[HTML]{F1F1F1} 53.2 (51.1) & {\cellcolor[HTML]{A50026}} \color[HTML]{F1F1F1} 74.6 (72.3) & {\cellcolor[HTML]{A50026}} \color[HTML]{F1F1F1} 15.0 (13.0) & {\cellcolor[HTML]{63BC62}} \color[HTML]{F1F1F1} 16.0 (13.2) & {\cellcolor[HTML]{A50026}} \color[HTML]{F1F1F1} 43.0 (38.5) & {\cellcolor[HTML]{4EB15D}} \color[HTML]{F1F1F1} 77.4 (87.7) & {\cellcolor[HTML]{A50026}} \color[HTML]{F1F1F1} 361.0 (330.0) \\
\addlinespace
\midrule
\addlinespace
\textbf{Joint} & \bfseries {\cellcolor{white}}  & \bfseries {\cellcolor{white}}  & \bfseries {\cellcolor{white}}  & \bfseries {\cellcolor{white}}  & \bfseries {\cellcolor{white}}  & \bfseries {\cellcolor{white}}  & \bfseries {\cellcolor{white}}  \\
$\mathrm{LEAR}$ & $-$ & $-$ & $-$ & $-$ & $-$ & {\cellcolor[HTML]{006837}} \color[HTML]{F1F1F1} 0.38 (0.37) & {\cellcolor[HTML]{006837}} \color[HTML]{F1F1F1} 2.50 (2.26) \\
$\mathrm{GAMLSS}$ & $-$ & $-$ & $-$ & $-$ & $-$ & {\cellcolor[HTML]{98D368}} \color[HTML]{000000} 0.42 (0.38) & {\cellcolor[HTML]{0C7F43}} \color[HTML]{F1F1F1} 2.56 (2.36) \\
$\mathrm{Naive}$ & $-$ & $-$ & $-$ & $-$ & $-$ & {\cellcolor[HTML]{A50026}} \color[HTML]{F1F1F1} 0.54 (0.54) & {\cellcolor[HTML]{A50026}} \color[HTML]{F1F1F1} 3.87 (3.51) \\
\bottomrule
\end{tabular}

%% file: opt_results_table_heatmap.tex
\begin{tabular}{lrrrrrrrr}
\toprule
 & \rotatebox[origin=l]{60}{\shortstack{Norm. costs\\(EUR)}} & \rotatebox[origin=l]{60}{\shortstack{Norm. energy\\(MWh)}} & \rotatebox[origin=l]{60}{\shortstack{Share abs.\\imb. energy}} & \rotatebox[origin=l]{60}{\shortstack{Imb. costs\\(EUR)}} & \rotatebox[origin=l]{60}{\shortstack{First violation\\(h)}} & \rotatebox[origin=l]{60}{\shortstack{Avg. contin.\\pump runtime\\(h)}} & \rotatebox[origin=l]{60}{\shortstack{Time\\(s)}} & \rotatebox[origin=l]{60}{\shortstack{Relative\\gap}} \\
\midrule
$\mathrm{FB\_DET\_FH}$ & {\cellcolor[HTML]{A50026}} \color[HTML]{F1F1F1} 6595.43 & \bfseries {\cellcolor[HTML]{006837}} \color[HTML]{F1F1F1} 74.73 & {\cellcolor[HTML]{FECC7B}} \color[HTML]{000000} 0.115 & {\cellcolor[HTML]{06733D}} \color[HTML]{F1F1F1} 0.00 & {\cellcolor[HTML]{18954F}} \color[HTML]{F1F1F1} 34.52 & \bfseries {\cellcolor[HTML]{006837}} \color[HTML]{F1F1F1} 15.08 & {\cellcolor[HTML]{108647}} \color[HTML]{F1F1F1} 45.47 & {\cellcolor[HTML]{006837}} \color[HTML]{F1F1F1} 0.01 \\
$\mathrm{Naive\_SH}$ & {\cellcolor[HTML]{FFF0A6}} \color[HTML]{000000} 6233.20 & {\cellcolor[HTML]{A50026}} \color[HTML]{F1F1F1} 75.71 & {\cellcolor[HTML]{E44C34}} \color[HTML]{F1F1F1} 0.153 & {\cellcolor[HTML]{E0F295}} \color[HTML]{000000} 170.78 & {\cellcolor[HTML]{FCA85E}} \color[HTML]{000000} 21.14 & {\cellcolor[HTML]{A50026}} \color[HTML]{F1F1F1} 7.15 & {\cellcolor[HTML]{FB9D59}} \color[HTML]{000000} 437.19 & {\cellcolor[HTML]{006837}} \color[HTML]{F1F1F1} 0.01 \\
$\mathrm{Naive\_DET\_FH}$ & {\cellcolor[HTML]{FAFDB8}} \color[HTML]{000000} 6181.58 & {\cellcolor[HTML]{FEEC9F}} \color[HTML]{000000} 75.28 & {\cellcolor[HTML]{A50026}} \color[HTML]{F1F1F1} 0.179 & {\cellcolor[HTML]{F16640}} \color[HTML]{F1F1F1} 340.24 & {\cellcolor[HTML]{A50026}} \color[HTML]{F1F1F1} 14.77 & {\cellcolor[HTML]{E0422F}} \color[HTML]{F1F1F1} 8.20 & {\cellcolor[HTML]{3CA959}} \color[HTML]{F1F1F1} 93.93 & {\cellcolor[HTML]{006837}} \color[HTML]{F1F1F1} 0.01 \\
$\mathrm{LEAR\_DET\_FH}$ & {\cellcolor[HTML]{E2F397}} \color[HTML]{000000} 6130.37 & {\cellcolor[HTML]{FDB163}} \color[HTML]{000000} 75.41 & {\cellcolor[HTML]{FCA85E}} \color[HTML]{000000} 0.127 & {\cellcolor[HTML]{A50026}} \color[HTML]{F1F1F1} 423.10 & {\cellcolor[HTML]{FA9656}} \color[HTML]{000000} 20.53 & {\cellcolor[HTML]{E24731}} \color[HTML]{F1F1F1} 8.25 & {\cellcolor[HTML]{96D268}} \color[HTML]{000000} 169.57 & {\cellcolor[HTML]{006837}} \color[HTML]{F1F1F1} 0.01 \\
$\mathrm{GAMLSS\_SH}^{**}$ & {\cellcolor[HTML]{E0F295}} \color[HTML]{000000} 6127.95 & {\cellcolor[HTML]{FDC574}} \color[HTML]{000000} 75.37 & {\cellcolor[HTML]{8CCD67}} \color[HTML]{000000} 0.047 & {\cellcolor[HTML]{8ECF67}} \color[HTML]{000000} 102.95 & {\cellcolor[HTML]{CFEB85}} \color[HTML]{000000} 28.28 & {\cellcolor[HTML]{F57245}} \color[HTML]{F1F1F1} 8.81 & {\cellcolor[HTML]{FFFBB8}} \color[HTML]{000000} 310.46 & {\cellcolor[HTML]{006837}} \color[HTML]{F1F1F1} 0.01 \\
$\mathrm{GAMLSS\_DET\_FH}$ & {\cellcolor[HTML]{D7EE8A}} \color[HTML]{000000} 6109.60 & {\cellcolor[HTML]{FEDE89}} \color[HTML]{000000} 75.32 & {\cellcolor[HTML]{FDB365}} \color[HTML]{000000} 0.124 & {\cellcolor[HTML]{CC2627}} \color[HTML]{F1F1F1} 388.57 & {\cellcolor[HTML]{FBA35C}} \color[HTML]{000000} 20.93 & {\cellcolor[HTML]{E65036}} \color[HTML]{F1F1F1} 8.37 & {\cellcolor[HTML]{FFF0A6}} \color[HTML]{000000} 331.39 & {\cellcolor[HTML]{006837}} \color[HTML]{F1F1F1} 0.01 \\
$\mathrm{GAMLSS\_SH}^{*}$ & {\cellcolor[HTML]{D3EC87}} \color[HTML]{000000} 6104.02 & {\cellcolor[HTML]{FCAA5F}} \color[HTML]{000000} 75.42 & {\cellcolor[HTML]{8ECF67}} \color[HTML]{000000} 0.048 & {\cellcolor[HTML]{A7D96B}} \color[HTML]{000000} 120.67 & {\cellcolor[HTML]{BDE379}} \color[HTML]{000000} 29.07 & {\cellcolor[HTML]{E95538}} \color[HTML]{F1F1F1} 8.43 & {\cellcolor[HTML]{FEEFA3}} \color[HTML]{000000} 335.05 & {\cellcolor[HTML]{006837}} \color[HTML]{F1F1F1} 0.01 \\
$\mathrm{GAMLSS\_SH}$ & {\cellcolor[HTML]{D3EC87}} \color[HTML]{000000} 6102.17 & {\cellcolor[HTML]{FCAA5F}} \color[HTML]{000000} 75.42 & {\cellcolor[HTML]{AFDD70}} \color[HTML]{000000} 0.058 & {\cellcolor[HTML]{ADDC6F}} \color[HTML]{000000} 125.22 & {\cellcolor[HTML]{BFE47A}} \color[HTML]{000000} 29.03 & {\cellcolor[HTML]{E44C34}} \color[HTML]{F1F1F1} 8.31 & {\cellcolor[HTML]{FFF8B4}} \color[HTML]{000000} 316.29 & {\cellcolor[HTML]{006837}} \color[HTML]{F1F1F1} 0.01 \\
$\mathrm{GAMLSS\_SH}_{\lambda^{\mathrm{risk}}=1}$ & {\cellcolor[HTML]{D1EC86}} \color[HTML]{000000} 6100.44 & {\cellcolor[HTML]{FB9D59}} \color[HTML]{000000} 75.44 & {\cellcolor[HTML]{AFDD70}} \color[HTML]{000000} 0.058 & {\cellcolor[HTML]{9BD469}} \color[HTML]{000000} 112.38 & {\cellcolor[HTML]{BBE278}} \color[HTML]{000000} 29.17 & {\cellcolor[HTML]{DD3D2D}} \color[HTML]{F1F1F1} 8.13 & {\cellcolor[HTML]{FFFAB6}} \color[HTML]{000000} 314.07 & {\cellcolor[HTML]{006837}} \color[HTML]{F1F1F1} 0.01 \\
$\mathrm{GAMLSS\_SH}_{\lambda^{\mathrm{risk}}=0}$ & {\cellcolor[HTML]{D1EC86}} \color[HTML]{000000} 6099.55 & {\cellcolor[HTML]{FCA55D}} \color[HTML]{000000} 75.43 & {\cellcolor[HTML]{ABDB6D}} \color[HTML]{000000} 0.056 & {\cellcolor[HTML]{A7D96B}} \color[HTML]{000000} 121.40 & {\cellcolor[HTML]{BDE379}} \color[HTML]{000000} 29.08 & {\cellcolor[HTML]{E44C34}} \color[HTML]{F1F1F1} 8.32 & {\cellcolor[HTML]{FDFEBC}} \color[HTML]{000000} 298.46 & {\cellcolor[HTML]{006837}} \color[HTML]{F1F1F1} 0.01 \\
$\mathrm{LEAR\_SH}$ & {\cellcolor[HTML]{CFEB85}} \color[HTML]{000000} 6097.34 & {\cellcolor[HTML]{DD3D2D}} \color[HTML]{F1F1F1} 75.59 & {\cellcolor[HTML]{D5ED88}} \color[HTML]{000000} 0.071 & {\cellcolor[HTML]{A0D669}} \color[HTML]{000000} 115.38 & {\cellcolor[HTML]{B7E075}} \color[HTML]{000000} 29.30 & {\cellcolor[HTML]{CE2827}} \color[HTML]{F1F1F1} 7.82 & {\cellcolor[HTML]{B3DF72}} \color[HTML]{000000} 199.10 & {\cellcolor[HTML]{006837}} \color[HTML]{F1F1F1} 0.01 \\
$\mathrm{GAMLSS\_SH}_{\widetilde{ED}_{0.5}}$ & {\cellcolor[HTML]{CBE982}} \color[HTML]{000000} 6091.17 & {\cellcolor[HTML]{FDB163}} \color[HTML]{000000} 75.41 & {\cellcolor[HTML]{A7D96B}} \color[HTML]{000000} 0.055 & {\cellcolor[HTML]{9DD569}} \color[HTML]{000000} 114.16 & {\cellcolor[HTML]{BDE379}} \color[HTML]{000000} 29.12 & {\cellcolor[HTML]{E24731}} \color[HTML]{F1F1F1} 8.25 & {\cellcolor[HTML]{FFFDBC}} \color[HTML]{000000} 307.27 & {\cellcolor[HTML]{006837}} \color[HTML]{F1F1F1} 0.01 \\
$\mathrm{GAMLSS\_SH}_{M^{\mathrm{red}}=10}$ & {\cellcolor[HTML]{C9E881}} \color[HTML]{000000} 6088.62 & {\cellcolor[HTML]{F57547}} \color[HTML]{F1F1F1} 75.50 & {\cellcolor[HTML]{BBE278}} \color[HTML]{000000} 0.062 & {\cellcolor[HTML]{89CC67}} \color[HTML]{000000} 99.58 & {\cellcolor[HTML]{9BD469}} \color[HTML]{000000} 30.42 & {\cellcolor[HTML]{DE402E}} \color[HTML]{F1F1F1} 8.15 & {\cellcolor[HTML]{A50026}} \color[HTML]{F1F1F1} 599.28 & {\cellcolor[HTML]{006837}} \color[HTML]{F1F1F1} 0.01 \\
$\mathrm{GAMLSS\_PF\_EPF\_SH}$ & {\cellcolor[HTML]{C1E57B}} \color[HTML]{000000} 6075.44 & {\cellcolor[HTML]{FDBB6C}} \color[HTML]{000000} 75.39 & {\cellcolor[HTML]{A5D86A}} \color[HTML]{000000} 0.054 & {\cellcolor[HTML]{98D368}} \color[HTML]{000000} 110.97 & {\cellcolor[HTML]{BFE47A}} \color[HTML]{000000} 29.00 & {\cellcolor[HTML]{E65036}} \color[HTML]{F1F1F1} 8.38 & {\cellcolor[HTML]{89CC67}} \color[HTML]{000000} 157.19 & {\cellcolor[HTML]{006837}} \color[HTML]{F1F1F1} 0.01 \\
$\mathrm{PF\_DET\_SH}$ & {\cellcolor[HTML]{2DA155}} \color[HTML]{F1F1F1} 5892.82 & {\cellcolor[HTML]{E8F59F}} \color[HTML]{000000} 75.16 & {\cellcolor[HTML]{0E8245}} \color[HTML]{F1F1F1} 0.011 & {\cellcolor[HTML]{108647}} \color[HTML]{F1F1F1} 16.52 & {\cellcolor[HTML]{18954F}} \color[HTML]{F1F1F1} 34.52 & {\cellcolor[HTML]{E54E35}} \color[HTML]{F1F1F1} 8.35 & \bfseries {\cellcolor[HTML]{006837}} \color[HTML]{F1F1F1} 6.61 & \bfseries {\cellcolor[HTML]{006837}} \color[HTML]{F1F1F1} 0.01 \\
$\mathrm{PF\_DET\_FH}$ & \bfseries {\cellcolor[HTML]{006837}} \color[HTML]{F1F1F1} 5790.76 & {\cellcolor[HTML]{E95538}} \color[HTML]{F1F1F1} 75.55 & \bfseries {\cellcolor[HTML]{006837}} \color[HTML]{F1F1F1} 0.001 & \bfseries {\cellcolor[HTML]{006837}} \color[HTML]{F1F1F1} -10.62 & \bfseries {\cellcolor[HTML]{006837}} \color[HTML]{F1F1F1} 36.64 & {\cellcolor[HTML]{EE613E}} \color[HTML]{F1F1F1} 8.60 & {\cellcolor[HTML]{1E9A51}} \color[HTML]{F1F1F1} 70.91 & {\cellcolor[HTML]{006837}} \color[HTML]{F1F1F1} 0.01 \\
\bottomrule
\end{tabular}